# Latest progress on the reduced-order particle-in-cell scheme: I. refining the underlying formulation

M. Reza*[1], F. Faraji*, A. Knoll*

* Plasma Propulsion Laboratory, Department of Aeronautics, Imperial College London, London, United Kingdom

**Abstract**: The particle-in-cell (PIC) method is a well-established and widely used kinetic plasma modelling approach that provides a hybrid Lagrangian-Eulerian approach to solve the plasma kinetic equation. Despite its power in capturing details of the underlying physics of plasmas, conventional PIC implementations are associated with a significant computational cost, rendering their applications for real-world plasma science and engineering challenges impractical. The acceleration of the PIC method has thus become a topic of high interest, with several approaches having been pursued to this end. Among these, the concept of reduced-order (RO) PIC simulations, first introduced in 2023, provides a uniquely flexible and computationally efficient framework for kinetic plasma modelling – characteristics that are extensively verified in various plasma configurations. In this two-part article, we report the latest progress achieved on RO-PIC. Part I article revisits the original RO-PIC formulation and introduces refinements that substantially enhance the cost-efficiency and accuracy of the method. We discuss these refinements in comparison against the original formulation, illustrating the progression to a "first-order" implementation from the baseline "zeroth-order" one. In a detailed step-by-step verification, we first test the newly updated reduced-dimension Poisson solver (RDPS) in the first-order RO-PIC against its zeroth-order counterpart using test-case Poisson problems. Next, comparing against the zeroth-order version, we examine the performance of the complete first-order RO-PIC code in two-dimensional plasma problems. One adopted plasma problem corresponds to electron plasma oscillations undergoing Landau damping, and the other to the Diocotron instability. The detailed verifications demonstrate that the improvements in the RO-PIC formulation enable the approach to provide full-2D-equivalent results at a substantially lower (up to an order of magnitude) computational cost compared to the zeroth-order RO-PIC.

## Section 1: Introduction

Existing computational tools and numerical simulations being used today to predict the plasma behavior and to study the involved physical processes differ significantly in terms of complexity, computational cost and fidelity. Fluid and hybrid fluid-kinetic simulations, while often computationally efficient, are limited by the lack of rigorous and generalizable closure models, which undermines their predictiveness. Despite numerous efforts over the years [1]-[5], closure models that are not reliant on a-priori assumptions or experimental tuning are yet to be realized [6]. Excellent progress has been recently made [7] to develop first-principles closure models that work well to explain some of the non-classical instability-induced effects that are not captured within conventional fluid/hybrid simulations. Nonetheless, even these models do not fully obviate the need for ad-hoc empirical data for practical modelling applications [7].

Particle-in-cell (PIC) simulations offer a first-principles approach to particle dynamics and are particularly well-suited for studying the kinetic behavior of complex plasmas such as those in E × B discharges, where plasma is immersed in perpendicular electric (E) and magnetic (B) fields. As PIC simulations provide high-fidelity, self-consistent predictions, they can in principle play a crucial role for understanding the performance and operation of advanced plasma technologies. However, the significant computational and memory resources required by conventional PIC codes, which is driven by their stringent numerical stability requirements, strongly limit the practicality of these kinetic models in real-world applications.

Recent techniques have been developed to alleviate the numerical constraints of traditional PIC simulations. Two key approaches include the Energy Conserving (EC) [8] and Direct Implicit (DI) PIC methods [9]. EC PIC allows for larger cell sizes by strictly conserving energy [8], although it sacrifices momentum conservation and is usually unsuitable for problems involving accelerated particle beams [10]. DI PIC, on the other hand, modifies the standard PIC algorithm to permit larger cell sizes and timesteps [9][11], though it requires careful selection of numerical parameters of the simulation and does not necessarily guarantee energy conservation [11][12]. The Sparse Grid (SG) method [13][14], a technique from computational mathematics, has been applied to PIC simulations [15][16] to reduce the number of grid cells, offering notable computational speed-ups, particularly for three-dimensional simulations [17]. All these methods provide different trade-offs between computational efficiency and accuracy in plasma modelling.

---

[1] **Corresponding Author** (m.reza20@imperial.ac.uk)

As a unique solution to address the pressing computational challenge of PIC simulations – distinct from the above works – we introduced in 2023 the reduced-order (RO) PIC scheme [18][19], which provides an approximation to the conventional PIC solution, resulting in a quasi-multi-dimensional representation of the problem at hand. The RO-PIC scheme relies on an innovative dimensionality-reduction treatment. It tackles multi-dimensional problems by breaking them down into a series of coupled one-dimensional problems, solving which are much more computationally affordable [20]. The number of these 1D problems controls the level of approximation, allowing for a flexible trade-off between simulation accuracy and computational cost [20].

The RO-PIC scheme has been extensively verified across various 2D plasma configurations [18][20][21]-[23]. The performed verifications entailed in-depth comparisons between the RO-PIC predictions and the results from conventional full-2D PIC simulations over a wide range of problem approximations determined by the number of "regions" used in RO-PIC (refer to Section 2 for the definition of regions). The rigorous assessments of the prediction fidelity of RO-PIC relative to its computational speedup were conducted in well-established benchmark cases, including the 2D axial-azimuthal Hall thruster benchmark problem of Ref. [24], the radial-azimuthal $E \times B$ benchmark case of Ref. [25], and the recently introduced Penning discharge benchmark described in Ref. [26]. Across all these cases, the RO-PIC scheme was shown to reproduce the full-2D results with an error of less than 10 % while achieving speedup factors ranging from 5 to 12. Notably, if we are willing to accept an error of up to 20 %, the speedup factor increase to 25. This highlights the RO-PIC's significant efficiency gains in computational performance while maintaining reasonable accuracy, making it a powerful tool for plasma simulations.

Accordingly, the RO-PIC has enabled conducting studies of both scientific and applied significance that were previously impractical with conventional PIC codes. These comprised: **(1)** unprecedentedly extensive high-fidelity parametric explorations of variations in plasma phenomena and behaviors across vast parameter spaces defined by electromagnetic field's strength and topology [27]-[30], as well as discharge parameters [31][32], **(2)** self-consistent kinetic simulations of a full-size 20kW-class industrial Hall thruster over time durations of real-world relevance [33].

In this work, we present the latest advancement on the RO-PIC, which involves revisiting the underlying formulation by improving the core ansatz. The original formulation of RO-PIC in Ref. [18] represents a "zeroth-order" approximation, whereas the revised formulation corresponds to a "first-order" approximation for multi-dimensional problems. The revised formulation leads to substantial improvements in both computational cost-efficiency and the overall accuracy characteristics of RO-PIC. As a result, the revisited RO-PIC can either achieve significantly greater computational speedups (beyond what was already offered by the original method) for the same accuracy level or deliver a significantly higher level of accuracy with the same computational resources. The original and revised RO-PIC formulations are discussed in Section 2 so as to establish how the legacy zeroth-order RO-PIC is transformed to the novel first-order one. This is then followed in Sections 3 and 4 by the presentation of verification results in several test cases and comparison of the respective predictions from zeroth-order and first-order RO-PIC implementations.

**Section 2: Description of the RO-PIC implementations in 2D**

The order-reduction or dimensionality-reduction technique that underlies RO-PIC is enabled through introducing a bespoke and innovative computational grid system, that we refer to as the "reduced-dimension" grid. The adoption of this grid system in PIC allows increasing the cell size beyond the Debye scale along one (for 2D) or two (for 3D) simulation dimension(s) at a time by an arbitrary amount, while maintaining the cell size along the other dimension(s) compatible with the requirements of the standard explicit momentum-conserving PIC method [34][35]. Accordingly, the number of computational cells – and consequently, the total number of macroparticles – decreases, leading to a reduction in the overall computational cost.

The schematic of the reduced-dimension (RD) grid in 2D, which we had also called a quasi-2D grid in Ref. [20], is illustrated in Figure 1. In this figure, subfigure (a) represents the RD grid system utilized in the original zeroth-order RO-PIC, whereas subfigure (b) presents the modified RD grid system underlying the revised first-order formulation.

In the RD grid system, the computational domain is divided into rectangular (in 2D, cubic in 3D) regions denoted by $\Omega$ using a "coarse" grid. In common between the zeroth-order and the first-order RO-PIC versions, each region $\Omega$ is then discretized independently along each involved simulation dimension using "fine, elongated" 1D cells. For the zeroth-order RO-PIC, these fine computational cells are placed within the regions (see the blue grids and the blue-shaded cells in Figure 1(a)). Whereas, in the first-order version, the fine 1D computational cells are modified to coincide with the boundaries of the regions (see the blue-shaded cells in Figure 1(b)).

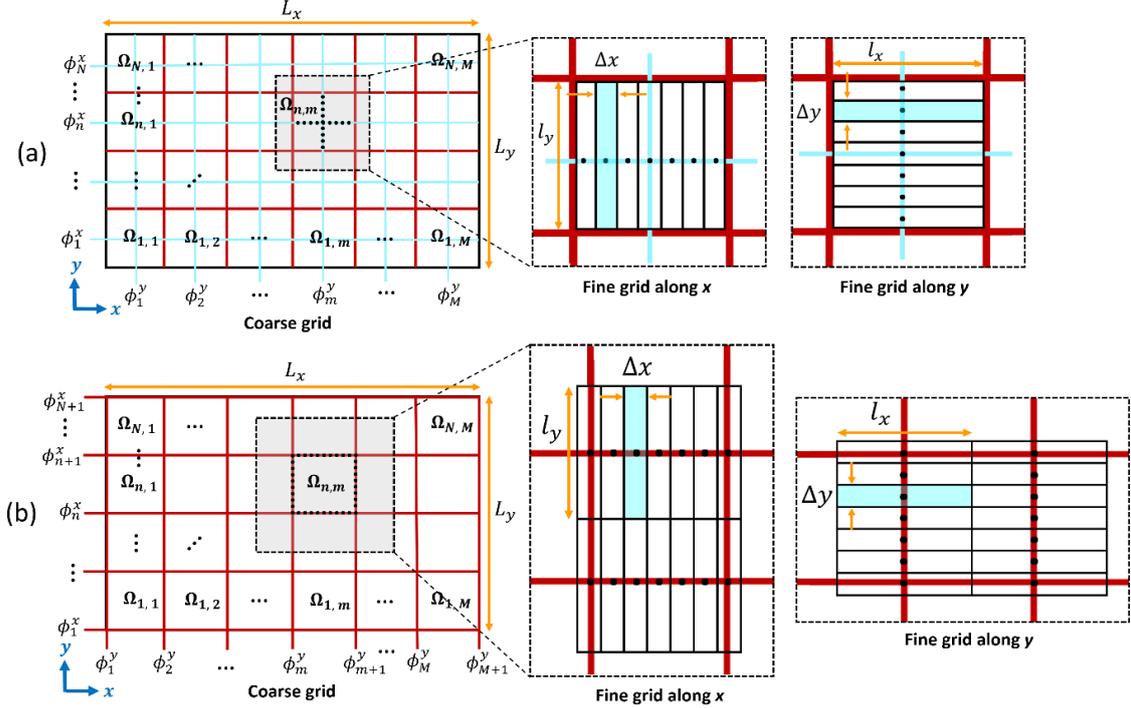

Figure 1: Schematic of the domain decomposition corresponding to the RD grid system for (a) zeroth-order RO-PIC and (b) first-order RO-PIC; (**left**) decomposition into multiple "regions" using a coarse grid, (**right**) the 1D computational cells (fine grids) for the discretization of each region along the $x$ and $y$ directions. In subfigure (a), the red lines represent the boundaries between the regions and the blue lines are the computational grids. In subfigure (b), the computational grids coincide with the regions' boundaries.

It is important to note that the cell sizes along each fine grid, i.e. $\Delta x$ and $\Delta y$, must be smaller than the Debye length in order to satisfy the stability requirements of explicit PIC schemes [35]. However, along the elongated side of the cells, the sizes $l_y$ and $l_x$ can be considerably larger than the Debye length. As pointed out in the beginning of this Section, this yields a substantial reduction in the total number of cells, thereby decreasing the overall number of required macroparticles.

The key difference between the zeroth-order and the first-order formulations lies in how, within each region, they treat the variation of grid-based data on the $x$-grid and the $y$-grid perpendicular to the respective grid directions. In the zeroth-order formulation, the data within each cell is assumed to remain constant across the entire elongated extent of the cell, i.e., over $l_y$ on the $x$-grid and over $l_x$ on the $y$-grid. In contrast, the first-order formulation assumes that the grid-based data vary linearly across the elongated extent of the cells. This means that, instead of being constant, the data changes smoothly from one side of the cells to the other, providing a higher level of accuracy compared to the zeroth-order method.

The RD grid system underpins "quasi"-multi-dimensional numerical simulations. Such a simulation can provide an approximation of multi-dimensional problems with arbitrary closeness to the corresponding "full"-multi-dimensional solution. It offers the flexibility to trade-off between prediction accuracy and computational cost as best suited for the specific purpose of the simulation. The degree of approximation of a reduced-order simulation relative to a full-2D/3D simulation – and thus its computational cost savings – is contingent upon the fineness of the coarse grid, which corresponds to the number of regions employed. In the 2D implementation of the reduced-order scheme, the total number of cells and the microparticles count scale with $O(N_i N_r)$ rather than $O(N_i^2)$, where $N_i$ and $N_r$ are the number of (fine) computational cells and the number of regions along one dimension, respectively. As a result, the aspect ratios of the fine cells – $l_x/\Delta x$ and $l_y/\Delta y$ – are directly proportional to the speedup factor achieved through the reduced-order scheme.

The incorporation of the RD grid and the associated dimensionality-reduction formulation – to be discussed in the subsequent Sections – into the PIC algorithm yields the RO-PIC scheme. The corresponding quasi-multi-dimensional PIC code is referred to as quasi-2D (Q2D) when accounting for two dimensions and as quasi-3D (Q3D) in a three-dimensional configuration.

The new PIC code developed in this research on the basis of the first-order formulation is called PICxelsior. It comprises a Q2D branch and a Q3D branch (part II [36]). Like previous PIC codes developed at Imperial Plasma Propulsion Laboratory (IPPL) and reported in Refs. [18][20]-[22], PICxelsior is written entirely in Julia language.

Our RO-PIC codes overall follow the standard algorithm of the explicit, momentum-conserving PIC method [34][35]. As discussed above, the key distinction between an RO-PIC and a conventional PIC lies in the computational grid. The grid is used for solving the electric potential and the electric field, as well as for depositing the particle-based data on the grid (handled by the "scatter" function) and gathering the grid-based data from the grid onto the particles' position (performed by the "gather" function). Therefore, the formulation and implementation of the functions and modules in RO-PIC that use or interact with the grid system must be adjusted. These include the scatter and the gather functions in addition to the Poisson solver. The following subsections provide details of these routines and modules for both the zeroth-order and the first-order RO-PIC

## 2.1. Scatter and gather functions

Figure 2 shows a schematic of the zeroth-order and the first-order RD grids. A generic particle $p$ is also shown within the computation domain as a sample to aid explain the scatter and gather functions.

The particle data, denoted by $\beta_p$, is deposited once onto the fine grid along the $x$-direction (the $x$-grid) and once onto the fine grid along $y$ (the $y$-grid) using Eq. 1.

$$\beta_n^x(x_i) = \sum_p \beta_p W_{n,i}^x(\mathbf{r}_i, \mathbf{r}_p), \qquad \beta_m^y(y_j) = \sum_p \beta_p W_{m,j}^y(\mathbf{r}_i, \mathbf{r}_p); \qquad \forall \mathbf{r}_i = (x_i, y_j) \in \Omega_{n,m}. \tag{Eq. 1}$$

Note that bold notations represent the position vectors. The summation over $p$ accounts for all the simulation particles. The average of the values on the two grids $x$ and $y$ will then reconstruct the distribution of the data on a full-2D grid using $\beta(x_i, y_j) = \frac{1}{2}\left(\beta_n^x(x_i) + \beta_m^y(y_j)\right)$.

Furthermore, we can gather grid-based data, $\beta^x$ and $\beta^y$, at a particle's location separately from the $x$-grid and the $y$-grid using Eq. 2.

$$\beta_p(x_p, y_p) = \frac{1}{2}\left(\sum_{n,i} \beta_n^x(x_i) W_{n,i}^x(\mathbf{r}_i, \mathbf{r}_p) + \sum_{m,j} \beta_m^y(y_j) W_{m,j}^y(\mathbf{r}_i, \mathbf{r}_p)\right); \qquad \forall \mathbf{r}_p = (x_p, y_p) \in \Omega_{n,m} \tag{Eq. 2}$$

where, $\beta_p$ is the gathered data at the particle's position, denoted by $x_p$ and $y_p$.

In the above relations, $W_{n,i}^x$ and $W_{m,j}^y$ are the interpolation weight functions that are dependent on the particle's position $\mathbf{r}_p$ relative to the neighboring grid nodes' location $\mathbf{r}_i$. These weight functions are defined in a manner consistent with the core assumption of either the zeroth-order or the first-order RO-PIC method as discussed earlier in Section 2 and will be further specified in subsection 2.2.

Accordingly, the weight functions for the zeroth-order RO-PIC scheme are provided by Eq. 3 and Eq. 4

$$W_{n,i}^x = \begin{cases} 1 - \left|\frac{x_p - x_i}{\Delta x}\right|, & \left|\frac{y_p - y_n}{l_y}\right| \leq 1 \text{ and } \left|\frac{x_p - x_i}{\Delta x}\right| \leq 1 \\ 0, & \left|\frac{y_p - y_n}{l_y}\right| > 1 \text{ or } \left|\frac{x_p - x_i}{\Delta x}\right| > 1 \end{cases}, \tag{Eq. 3}$$

$$W_{m,j}^y = \begin{cases} 1 - \left|\frac{y_p - y_j}{\Delta y}\right|, & \left|\frac{x_p - x_m}{l_x}\right| \leq 1 \text{ and } \left|\frac{y_p - y_j}{\Delta y}\right| \leq 1 \\ 0, & \left|\frac{x_p - x_m}{l_x}\right| > 1 \text{ or } \left|\frac{y_p - y_j}{\Delta y}\right| > 1 \end{cases}, \tag{Eq. 4}$$

and, for the first-order RO-PIC are given by Eq. 5 and Eq. 6

$$W_{n,i}^x = \begin{cases} \left(1 - \left|\frac{y_p - y_n}{l_y}\right|\right)\left(1 - \left|\frac{x_p - x_i}{\Delta x}\right|\right), & \left|\frac{y_p - y_n}{l_y}\right| \leq 1 \text{ and } \left|\frac{x_p - x_i}{\Delta x}\right| \leq 1 \\ 0, & \left|\frac{y_p - y_n}{l_y}\right| > 1 \text{ or } \left|\frac{x_p - x_i}{\Delta x}\right| > 1 \end{cases}, \tag{Eq. 5}$$

$$W_{m,j}^{y} = \begin{cases} \left(1 - \left|\frac{x_p - x_m}{l_x}\right|\right)\left(1 - \left|\frac{y_p - y_j}{\Delta y}\right|\right), & \left|\frac{x_p - x_m}{l_x}\right| \leq 1 \text{ and } \left|\frac{y_p - y_j}{\Delta y}\right| \leq 1 \\ 0, & \left|\frac{x_p - x_m}{l_x}\right| > 1 \text{ or } \left|\frac{y_p - y_j}{\Delta y}\right| > 1 \end{cases}.$$ (Eq. 6)

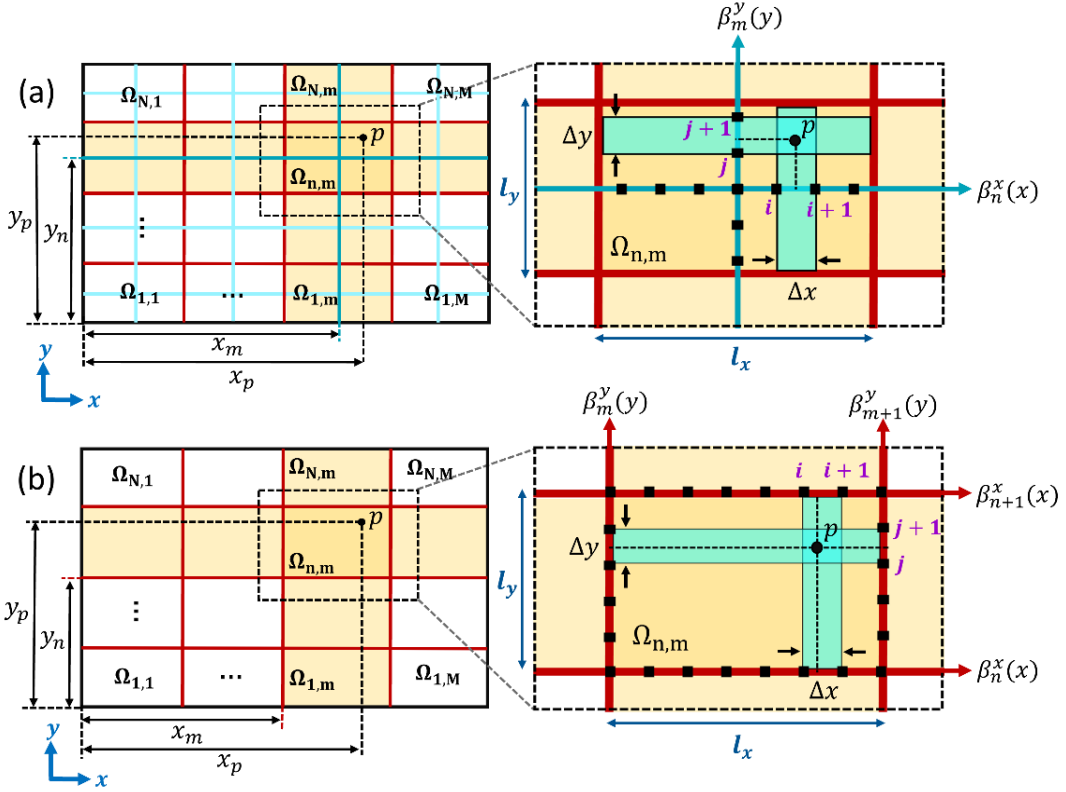

Figure 2: Schematic illustration of the exchange of data between the particles and the RD grid in (a) zeroth-order, and (b) first-order RO-PIC.

### 2.2. Poisson solver

The discretization of the differential equations for solving the electromagnetic fields must be modified in RO-PIC for compatibility with the RD grid configuration.

In this work, we focus on the electrostatic PIC method, which relies on solving the Poisson's/Gauss's equation to obtain the electric potential. Accordingly, we derive in the following the formulation of the reduced-dimension (quasi-2D) Poisson's equation that will be in turn incorporated within the electrostatic (Q2D) RO-PIC.

We start by applying Gauss's law ($\int_{\partial A} \mathbf{E}.dl = 1/\epsilon_0 \int_A \rho dA$) to the cells along the $x$ and the $y$ grids, as shown in Figure 3, substituting $\mathbf{E} = -\nabla \phi$ and dividing by the respective cell sizes, $\Delta x$ and $\Delta y$, to obtain Eqs. 7 and 8

$$\left(\frac{\partial \phi}{\partial y}\Big|_{y_N} - \frac{\partial \phi}{\partial y}\Big|_{y_S}\right) + \frac{1}{\Delta x}\int_{y_S}^{y_N}\left(\frac{\partial \phi}{\partial x}\Big|_{x_E} - \frac{\partial \phi}{\partial x}\Big|_{x_W}\right)dy = \frac{1}{\epsilon_0}\int_{y_S}^{y_N}\rho dy,$$ (Eq. 7)

$$\left(\frac{\partial \phi}{\partial x}\Big|_{x_E} - \frac{\partial \phi}{\partial x}\Big|_{x_W}\right) + \frac{1}{\Delta y}\int_{x_W}^{x_E}\left(\frac{\partial \phi}{\partial y}\Big|_{y_N} - \frac{\partial \phi}{\partial y}\Big|_{y_S}\right)dx = \frac{1}{\epsilon_0}\int_{x_W}^{x_E}\rho dx,$$ (Eq. 8)

where $\rho$ is the charge density distribution.

We now need to substitute the core ansatz of RO-PIC for the solution of $\phi$ in Eqs. 7 and 8. For clarity, we first do this to derive the zeroth-order formulation, followed by the presentation of how the first-order equations are obtained.

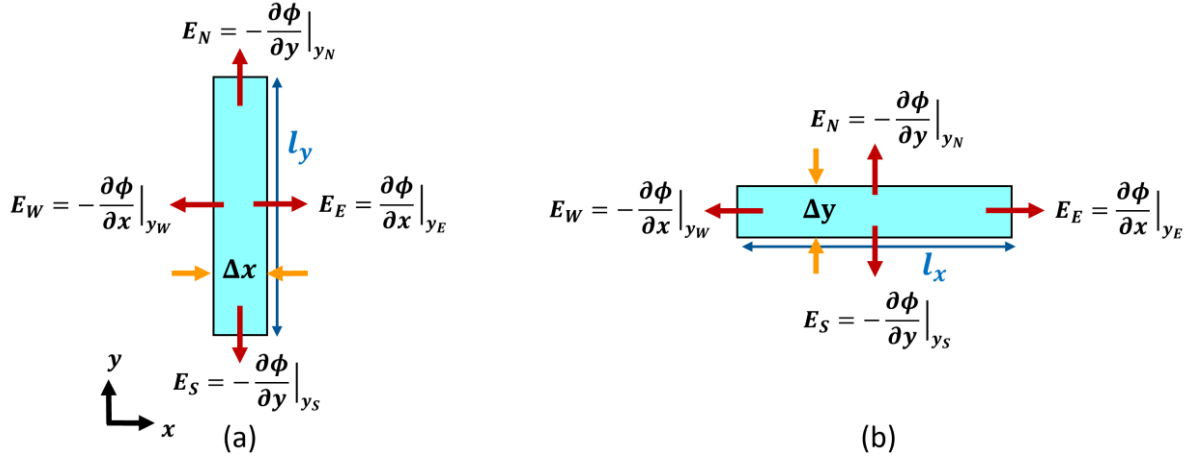

Figure 3: Schematic of the computational elements in region $\Omega_{n,m}$ associated with the RO-PIC's RD grid, showing the electric field components across the elements' sides; (a) element along the $x$-grid, and (b) element along the $y$-grid. N in this figure stands for North, S for South, E for East and W for West.

A) **Zeroth-order quasi-2D Poisson's equations**: The zeroth-order Q2D approximation to the Poisson's equation is grounded in the core ansatz that, within each region $\Omega_{n,m}$ (where $m = 1,2,\ldots M$ and $n = 1,2,\ldots,N$), the two-dimensional distribution of the electric potential ($\phi$) is approximated as the direct superposition of potential functions $\phi_x$ and $\phi_y$

$$\phi(x,y) \approx \phi_n^x(x) + \phi_m^y(y), \qquad \forall x, y \in \Omega_{n,m}, \tag{Eq. 9}$$

In Eq. 9, $\phi_n^x$ is assumed to remain constant across the vertical extent ($l_y$) of region $\Omega_{n,m}$, while $\phi_m^y$ is constant across the horizontal extent ($l_x$) of the region $\Omega_{n,m}$. Substituting Eq. 9 in Eqs. 7 and 8 yields

$$\begin{aligned}
l_y \frac{d^2\phi_n^x}{dx^2} &+ \frac{1}{\Delta x}\int_{y_S}^{y_N}\left(\frac{\partial \phi_m^y}{\partial x}\bigg|_{x_E} - \frac{\partial \phi_m^y}{\partial x}\bigg|_{x_W}\right)dy + \left(\frac{\partial \phi_m^y}{\partial y}\bigg|_{y_N} - \frac{\partial \phi_m^y}{\partial y}\bigg|_{y_S}\right) \\
&+ \left(\frac{\partial \phi_n^x}{\partial y}\bigg|_{y_N} - \frac{\partial \phi_n^x}{\partial y}\bigg|_{y_S}\right) = -\frac{1}{\epsilon_0}\int_{y_S}^{y_N}\rho dy,
\end{aligned} \tag{Eq. 10}$$

$$\begin{aligned}
l_x \frac{d^2\phi_m^y}{dy^2} &+ \frac{1}{\Delta y}\int_{x_W}^{x_E}\left(\frac{\partial \phi_n^x}{\partial y}\bigg|_{y_N} - \frac{\partial \phi_n^x}{\partial y}\bigg|_{y_S}\right)dx + \left(\frac{\partial \phi_n^x}{\partial x}\bigg|_{x_E} - \frac{\partial \phi_n^x}{\partial x}\bigg|_{x_W}\right) \\
&+ \left(\frac{\partial \phi_m^y}{\partial x}\bigg|_{x_E} - \frac{\partial \phi_m^y}{\partial x}\bigg|_{x_W}\right) = -\frac{1}{\epsilon_0}\int_{x_W}^{x_E}\rho dx.
\end{aligned} \tag{Eq. 11}$$

B) **First-order quasi-2D Poisson's equations**: In the first-order approximation, the potential functions $\phi_x$ and $\phi_y$ are assumed to vary linearly across their respective regions. In other words, at any given position within region $\Omega_{n,m}$, the two-dimensional electric potential is approximated as the linear superposition of $\phi_x$ and $\phi_y$ calculated on the boundaries of the regions, resulting in the following modification to Eq. 9

$$\phi(x,y) \approx (1-\tilde{y})\phi_n^x(x) + \tilde{y}\phi_{n+1}^x(x) + (1-\tilde{x})\phi_m^y(y) + \tilde{x}\phi_{m+1}^y(y), \quad \forall \tilde{x}, \tilde{y} \in \Omega_{n,m}, \tag{Eq. 12}$$

where, $\tilde{x} = \frac{x-x_m}{l_x}$ and $\tilde{y} = \frac{y-y_n}{l_y}$ are linear interpolation weights, which are the normalized relative coordinates within the region.

Replacing Eq. 12 in Eqs. 7 and 8 leads to

$$\begin{aligned}
\frac{l_y}{8}\left(\frac{d^2\phi_{n-1}^x}{dx^2} + 6\frac{d^2\phi_n^x}{dx^2} + \frac{d^2\phi_{n+1}^x}{dx^2}\right) &+ (1-\tilde{x})\left(\frac{\partial \phi_m^y}{\partial y}\bigg|_{y_N} - \frac{\partial \phi_m^y}{\partial y}\bigg|_{y_S}\right) \\
&+ \tilde{x}\left(\frac{\partial \phi_{m+1}^y}{\partial y}\bigg|_{y_N} - \frac{\partial \phi_{m+1}^y}{\partial y}\bigg|_{y_S}\right) + \frac{1}{l_y}(\phi_{n-1}^x - 2\phi_n^x + \phi_{n+1}^x) \\
&+ \frac{1}{l_x}\int_{y_S}^{y_N}(\phi_{m+2}^y - 2\phi_{m+1}^y + \phi_m^y)dy = -\frac{1}{\epsilon_0}\int_{y_S}^{y_N}\rho dy
\end{aligned} \tag{Eq. 13}$$

$$\frac{l_x}{8}\left(\frac{d^2\phi^y_{m-1}}{dy^2} + 6\frac{d^2\phi^y_m}{dy^2} + \frac{d^2\phi^y_{m+1}}{dy^2}\right) + (1-\tilde{y})\left(\frac{\partial\phi^x_n}{\partial x}\bigg|_{x_E} - \frac{\partial\phi^x_n}{\partial x}\bigg|_{x_W}\right)$$
$$+ \tilde{y}\left(\frac{\partial\phi^x_{n+1}}{\partial x}\bigg|_{x_E} - \frac{\partial\phi^x_{n+1}}{\partial x}\bigg|_{x_W}\right) + \frac{1}{l_x}(\phi^y_{m-1} - 2\phi^y_m + \phi^y_{m+1}) \quad \text{(Eq. 14)}$$
$$+ \frac{1}{l_y}\int_{x_W}^{x_E}(\phi^x_{n+2} - 2\phi^x_{n+1} + \phi^x_n)dx = -\frac{1}{\epsilon_0}\int_{x_W}^{x_E}\rho\, dx.$$

The reduced-dimension Poisson's system of coupled differential equations described by either Eqs. 10 and 11 for the zeroth-order approximation or by Eqs. 13 and 14 for the first-order approximation are solved numerically. The PICxelsior features a bespoke electric potential field solver, which we have termed reduced-dimension Poisson solver or RDPS [18]-[20]. The RDPS numerically solves either of the above systems of equations using Julia's built-in direct matrix solve algorithm based on the LU decomposition method.

**Section 3: Standalone verifications of the RDPS**

We compare the approximate Q2D solutions as provided by RDPS at various number-of-regions against the corresponding full-2D solutions to the Poisson's equation. This comparison is conducted for several test cases featuring different source terms and boundary conditions to assess the convergence behavior of RDPS for a wide range of gradients involved in the problem. The definition of the source terms ($\rho(x,y)$) and the boundary conditions are presented in Table 1.

We would underline that the solution in case 1 is dominated by the boundary condition, whereas in case 2, the solution is primarily influenced by the charge distribution. In either case, various wavenumbers ($n$) indicate different levels of gradient in the resulting potential distribution.

In all cases, the domain is a Cartesian ($x - y$) plane of unit length along the $x$ and $y$ directions. For simplicity, the permittivity of free space ($\epsilon_0$) is assumed to be unity for all problems. The number of computational cells along both the $x$ and $y$ directions are $N_i = N_j = 100$ for all tests performed.

| Case No. | Source terms ($\rho(x,y)$) | Boundary Conditions |
|---|---|---|
| 1 | $\rho = 0$ | $\phi(0,y) = \phi(L_x, y) = \sin(n\pi y);\ n = 1,2,3,4,5$ <br> $\phi(x,0) = \phi(x, L_y) = sin(n\pi y);\ n = 1,2,3,4,5$ |
| 2 | $\rho = \sin(2\pi r)\cos(n\theta)\exp(-5r^2);\ n = 2,3,4,5,6$ <br> $r = \sqrt{(x-0.5)^2 + (y-0.5)^2},$ <br> $\theta = \tan^{-1}\left(\frac{y-0.5}{x-0.5}\right).$ | $\phi(0,y) = \phi(L_x, y) = 0$ <br> $\phi(x,0) = \phi(x, L_y) = 0$ |

Table 1: Definition of the 2D Poisson problems for standalone verification of the reduced-dimension Poisson's solver.

As a first set of results, the solutions of the zeroth-order and the first-order RDPS, using different number-of-regions, are compared for case 1 with $n = 1$ and for case 2 with $n = 2$ in Figure 4(a) and (b), respectively.

These results clearly demonstrate that the first-order formulation yields a remarkable improvement, enhancing both the smoothness of the solution and its overall accuracy in comparison to the full-2D solution. The improvement is particularly significant for lower number-of-regions ($M$ and $N \leq 10$ ). In particular, the smoothness of the zeroth-order solution with 25 regions is comparable to that of the first-order solution with only 3 or 5 regions. This observation suggests an additional fivefold speedup that the first-order RO-PIC could provide on top of that already achieved via the zeroth-order RO-PIC. This will be further demonstrated for the plasma test case 2, to be discussed in subsection 4.2.2.

The first-order solutions from RDPS for case 1 with $n = 2, 3, 4$ and 5, and for case 2 with $n = 3, 4, 5$ and 6 are presented in Figure 5 and Figure 6, respectively.

It is noticed from these figures that, at the low number-of-regions limit, i.e., $M, N \leq 5$, the first-order RDPS successfully provides a reasonable approximation to the full-2D results. At the intermediate number-of-region ($M = N = 10$), the first-order RDPS' solutions are almost indistinguishable from the full-2D ones from the visual qualitative point of view.

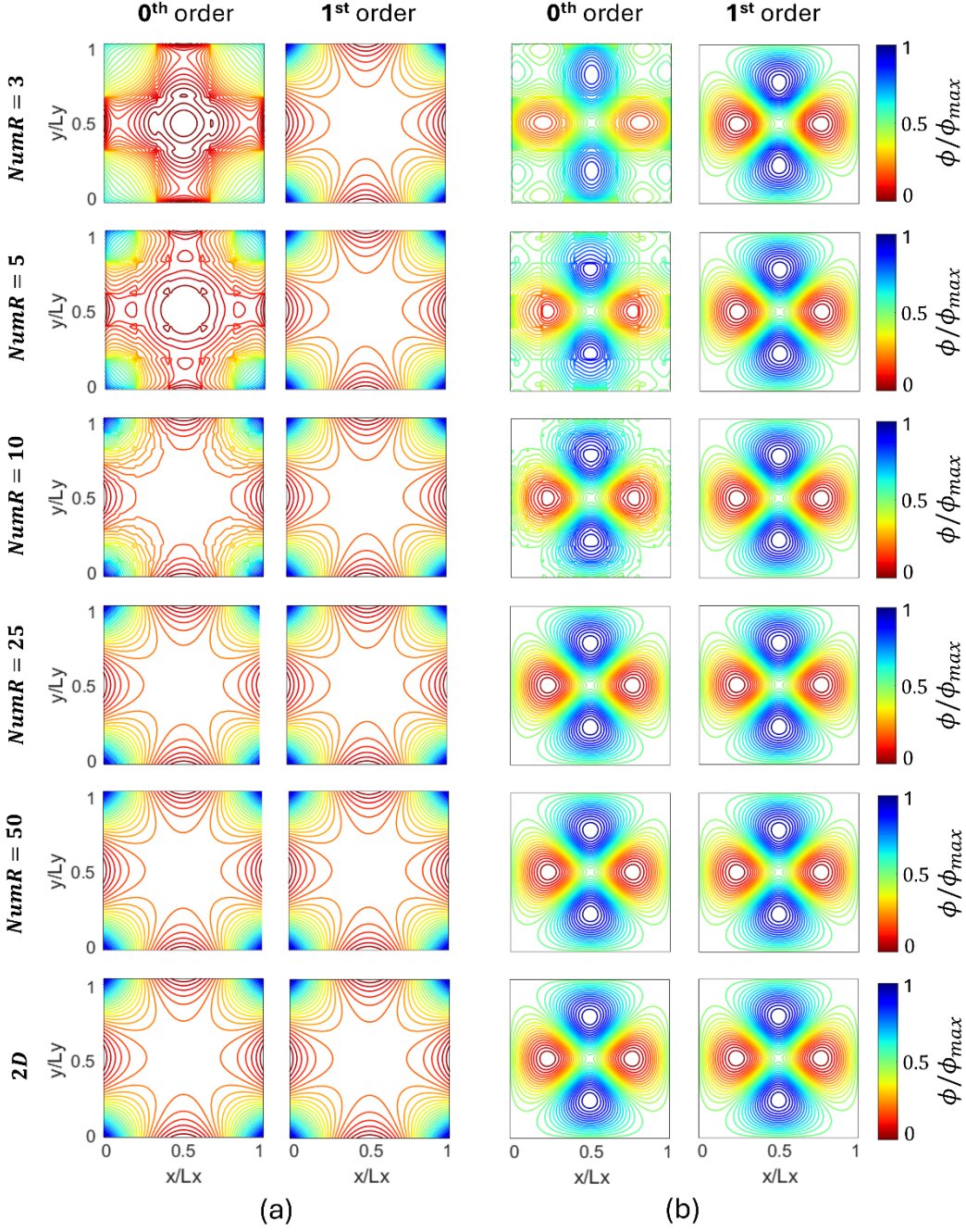

Figure 4: Comparison of the zeroth-order and the first-order RDPS' solutions with various number of regions against the corresponding full-2D solution (**bottom-most row**) for (a) case 1 with $n = 1$ and (b) case 2 with $n = 2$. The numbers of regions are equal along both dimensions ($M = N$), which are denoted by *NumR*.

In order to establish a quantitative framework of assessment, the error convergence characteristics of the RDPS' solutions for the studied test cases are provided in Figure 7.

In the plots of this figure, the errors of the zeroth-order and the first-order RDPS are compared as the number-of-regions increases. The error term has been defined by Eq. 15 as the normalized L2 norm of the difference between the Q2D approximation ($\phi_{Q2D}$) from RDPS with different number of regions and the full-2D solution ($\phi_{2D}$).

$$Error = 100 \frac{\sqrt{\sum_{i=1}^{N_i} \sum_{j=1}^{N_j} \left(\phi_{2D}(i,j) - \phi_{Q2D}(i,j)\right)^2}}{\sqrt{\sum_{i=1}^{N_i} \sum_{j=1}^{N_j} \left(\phi_{2D}(i,j)\right)^2}}.$$  (Eq. 15)

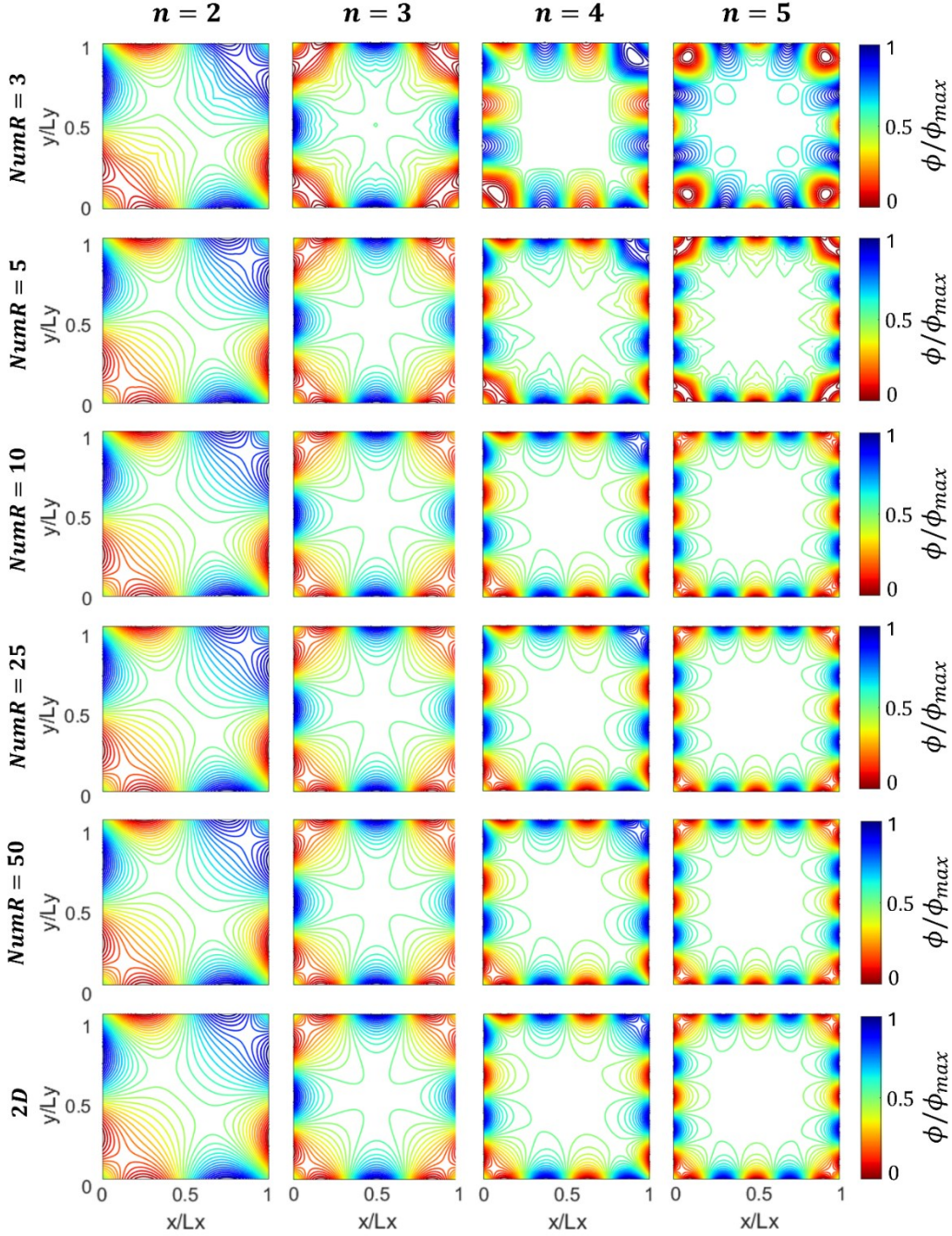

Figure 5: Comparison of the first-order RDPS' solutions with various number of regions against the corresponding full-2D solution (**bottom-most row**) for case 1 with different mode numbers (n). The numbers of regions are equal along both dimensions ($M = N$), which are denoted by *NumR*.

The plots of Figure 7 illustrate that the error in the first-order RDPS drops much faster than in the zeroth-order RDPS. As a result, the first-order RDPS can achieve over an order of magnitude greater accuracy than the zeroth-order solver for the same number of regions. This advancement allows us to achieve the same level of accuracy with significantly fewer regions (coarser grid), thereby further reducing computational costs. In this regard, the ratio of the number of cells used in the RDPS ($N_{cell,Q2D}$) to those in a full-2D Poisson solver ($N_{cell,2D}$), displayed on the plots of Figure 7, is directly correlated with the corresponding ratio of computational times between the RDPS and the 2D Poisson solver.

We would also point out that the RDPS' convergence behavior is seen from Figure 7 to vary depending on the specific case and the significance of the involved gradients. However, even in the presence of notably steep gradients in case 1 with $n = 5$ and in case 2 with $n = 6$, the errors remain below 2 % and 5 %, respectively, when using 10 regions.

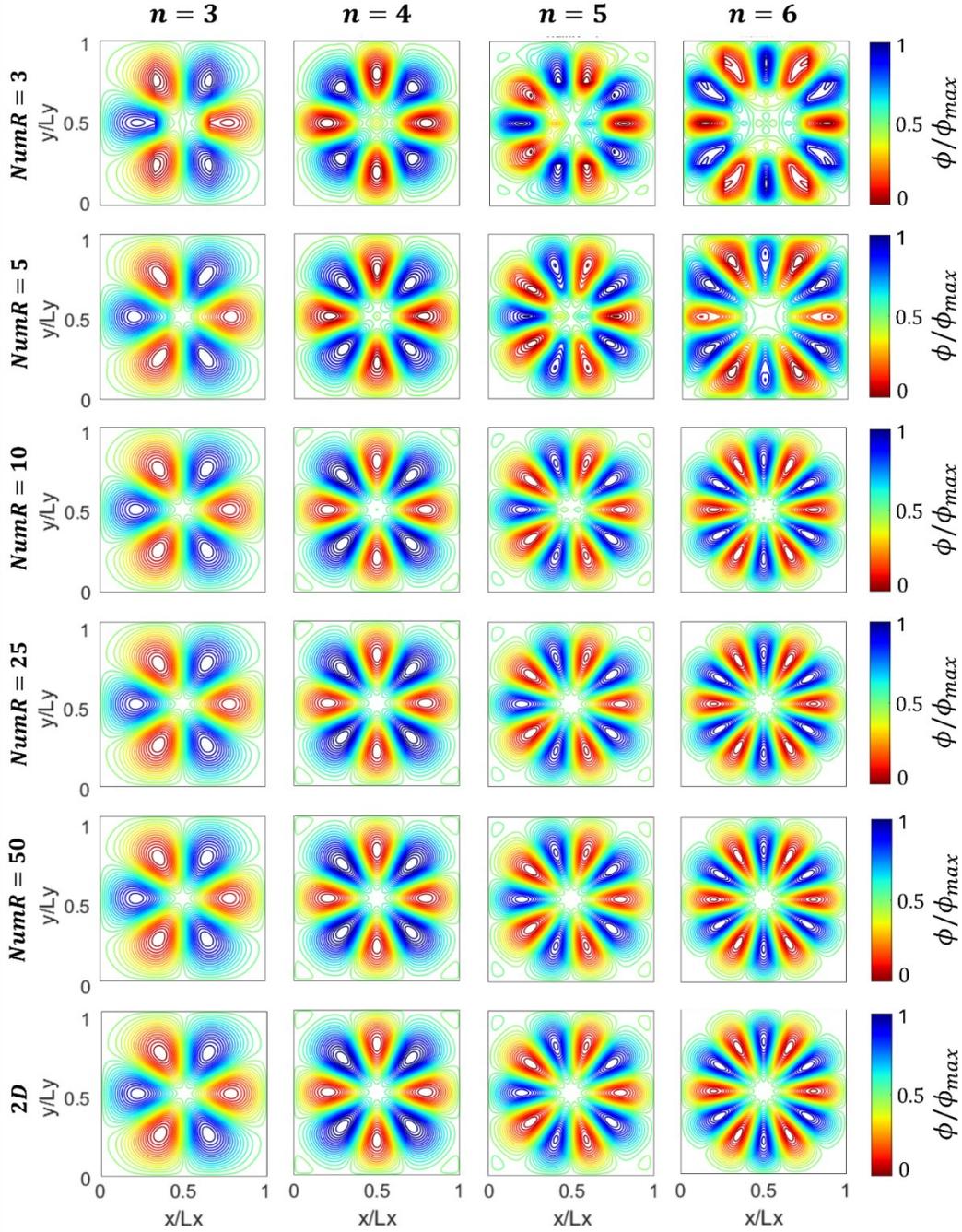

Figure 6: Comparison of the first-order RDPS' solutions with various number of regions against the corresponding full-2D solution (**bottom-most row**) for case 2 with different mode numbers (n). The numbers of regions are equal along both dimensions ($M = N$), which are denoted by *NumR*.

Before moving to the next Section, we highlight that additional comparisons between the first-order and the zeroth-order RDPS for a wider variety of test cases, including Poisson problems with periodic or Neumann boundary conditions, are presented in Appendix A. The results in Appendix A demonstrate similar improvements in the first-order solutions compared to the zeroth-order ones, reinforcing the arguments and the conclusions provided in the Section.

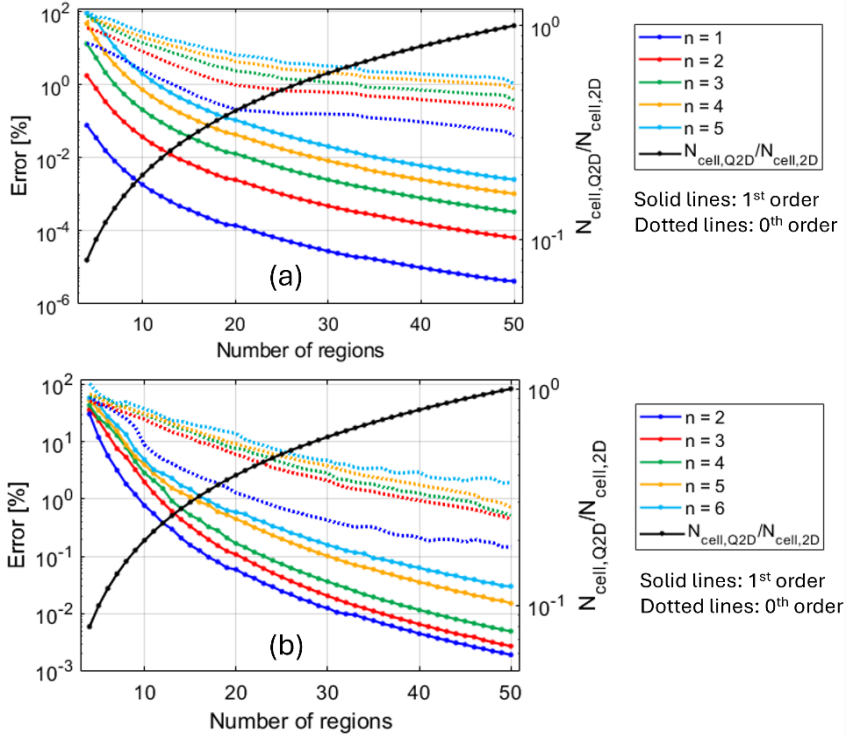

Figure 7: Error convergence of the RDPS' solution; variation of the normalized L2 error (Eq. 15) between the full-2D solution and the approximate Q2D solution from the RDPS when increasing number of regions for (a) case 1 and (b) case 2. The solid lines and the dotted lines represent, respectively, the error of the solutions from the first-order and zeroth-order formulations. The right axis represents the ratios of cells in Q2D to that in the full-2D for the respective number-of-regions. The y-axes are in logarithmic scale.

**Section 4: RO-PIC verifications**

We now aim to verify the first-order RO-PIC code in representative plasma test cases. To this end, we present the results the first-order Q2D PIC simulations for two plasma problems: **(1)** the problem of electron plasma oscillations undergoing Landau damping [37], and **(2)** the problem of Diocotron instability [38][39]. For test case 2, we also compare the performance of the first-order RO-PIC against that of the zeroth-order version.

For all the Q2D simulations discussed in this Section, equal numbers of regions with uniform extents are used along the $x$ and the $y$ directions of the computational domain. The reference full-2D results, against which the outcomes of the Q2D simulations are compared, are obtained using the IPPL-2D PIC code, which is itself verified extensively in Refs. [20][23].

**4.1. Test case 1: Electron plasma oscillations**

*4.1.1. Description of the problem setup*

The computational domain is a 2D Cartesian $(x-y)$ plane, as illustrated in Figure 8, with equal dimensions of $2\ cm$ along both simulation axes.

Initially, electrons are sampled from a Maxwellian distribution function with an electron temperature $(T_{e,0})$ of $0.1\ eV$ and loaded into a quarter of the domain, specifically within the region $x \in [0, L_x/2)$, $y \in [0, L_y/2)$, with a uniform density of $n_0 = 1 \times 10^{11}\ m^{-3}$. The ions are loaded uniformly over the entire domain with the same density as the electrons $(n_0)$ and no initial velocity $(T_{i,0} = 0)$. The ions used in this simulation are oxygen ions (O+), and their positions and velocities are continuously updated throughout the simulation.

All four boundaries are grounded, with a zero-volt Dirichlet condition applied for the potential solver. For the particle boundary conditions, any particle that reaches the boundaries is specularly reflected back into the domain. The physical and numerical parameters used in the simulations are provided in Table 2. This test case is inspired by the example problem in Ref. [37].

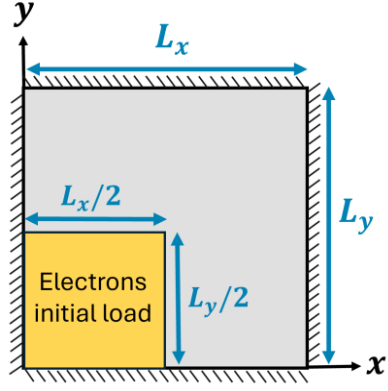

Figure 8: Schematic of the computational domain and setup of test case 1.

The imposed charge separation in this problem represents an electrostatic perturbation driving plasma oscillations, which are damped over time by Landau damping [40]. Landau damping is a non-collisional mechanism which damps the plasma oscillations due to resonant interactions between the oscillating electric field and electrons in the plasma. When the phase velocity of the plasma wave ($v_{ph}$) is close to the velocities of electrons in the tail of velocity distribution function, a resonant energy exchange occurs. Electrons with velocities slightly below $v_{ph}$ gain energy from the wave. The net effect is a transfer of energy from the wave to the particles, leading to a gradual decay in the wave's amplitude.

| Parameter | Value [unit] |
|---|---|
| Physical Parameters | |
| Initial plasma density ($n_0$) | $1 \times 10^{11}$ [m$^{-3}$] |
| Initial electron temperature ($T_{e,0}$) | 0.1 [eV] |
| Initial ion temperature ($T_{i,0}$) | 0.0 [eV] |
| Potential at walls ($\phi_w$) | 0 [V] |
| Ion mass | $2.65686 \times 10^{-26}$ [kg] |
| Computational Parameters | |
| Time step ($\Delta t$) | $1 \times 10^{-9}$ [s] |
| Time averaging for outputs | $1 \times 10^{-8}$ [s] |
| Total simulation time ($t_{sim}$) | 30 [µs] |
| Domain length ($L_x = L_y$) | 2 [cm] |
| Cell size ($\Delta x = \Delta y$) | $5 \times 10^{-2}$ [cm] |
| Number of computational nodes along each direction ($N_i = N_j$) | 40 |
| Initial number of macroparticles per cells ($N_{ppc}$) | 50 |

Table 2: Summary of the main computational and physical parameters used for simulations of test case 1.

### 4.1.2. Results

The electron plasma oscillations and their rate of damping can be observed in the evolution of the electric potential energy ($E_{pot}$) of the system and the kinetic energy ($E_{kin}$) of the particles (Figure 9). These quantities are obtained from Eq. 16 and Eq. 17, respectively

$$E_{pot} = \frac{1}{\epsilon_0} \sum_{i=1}^{N} |E_i|^2 V_i,  \quad\quad\quad (\text{Eq. 16})$$

$$E_{kin} = \frac{1}{2} \sum_{k=1}^{K} W_k m_k |v_k|^2. \quad\quad\quad (\text{Eq. 17})$$

In Eq. 16, $|E_i|$ represents the magnitude of the electric field in each cell, $V_i$ is the volume of the computational cell, and the summation is taken over total number of cells ($N$). In Eq. 17, $|v_k|$ denotes the magnitude of particle's velocity, $m_k$ is the particle's mass, $W_k$ is the macroparticle's weight, with the summation encompassing all macroparticles in the system, including both ions and electrons.

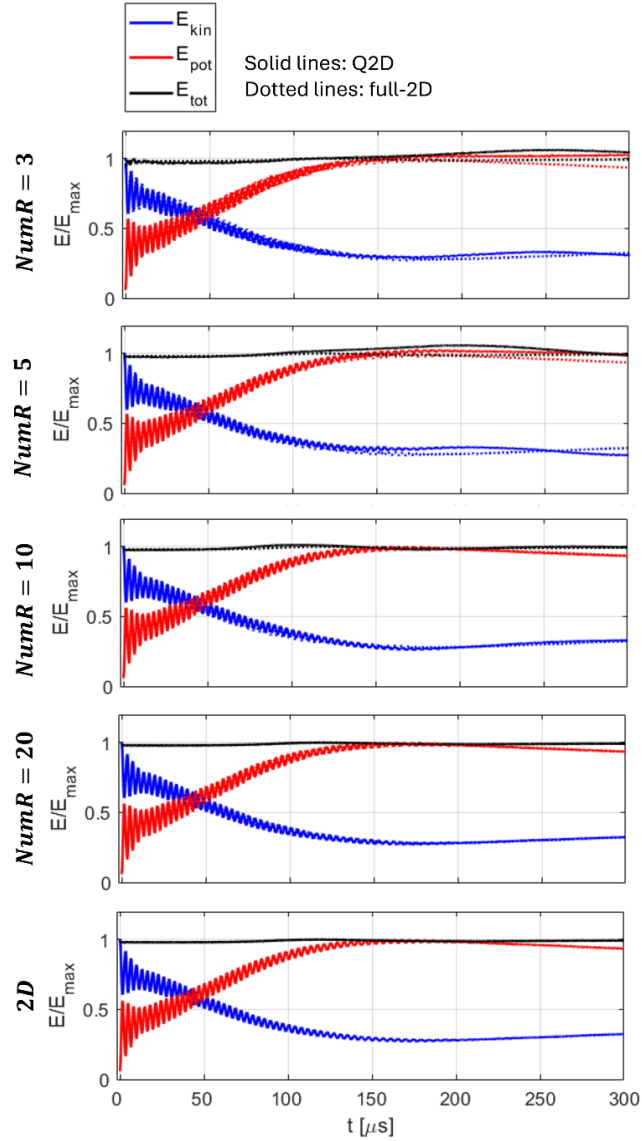

Figure 9: Time evolution of the total electric potential energy ($E_{pot}$), the total electrons' kinetic energy ($E_{kin}$), and the total energy of the system ($E_{tot} = E_{pot} + E_{kin}$) from the first-order Q2D and the full-2D simulations for test case 1. The results from the full-2D simulations are superimposed as dotted lines on the Q2D plots. The normalization factor ($E_{max}$) is the maximum of the respective energies (potential and kinetic) from the full-2D simulation over the simulated duration.

Looking at the plots in Figure 9, we find that the evolution of the potential and kinetic energies captured in the Q2D simulations closely aligns with the corresponding results from the full-2D simulation. Additionally, the total energy of the system is shown to remain nearly constant across the simulations, with a maximum variation of less than 5% observed in the Q2D simulation with low number-of-regions ($M, N = 3$ and $5$).

Figure 10 presents three sample snapshots of the resolved axial electric field ($E_x$) at different time instants through the simulations. These figures provide an illustrative comparison of how effectively each approximation of the Q2D simulations captures the evolving electric field during the oscillations. In particular, the simulation with $M = N = 20$ regions closely replicates the full-2D results, accurately capturing the detailed structure of the evolving electric field. Simulations with fewer regions still resolve the essential features of the oscillations, though they exhibit minor differences in finer details, reflecting the trade-off between computational efficiency and accuracy.

Finally, to provide a quantitative measure of accuracy, we have presented in Figure 11 the errors associated with the characteristics of the oscillations resolved in the Q2D simulations. In particular, the errors are related to the oscillation frequency and the average damping rate, as observed in the variations of electric potential and particles' kinetic energies (Figure 9). The damping rates are calculated over the time interval of $0 - 100\ \mu s$. The frequency of the oscillations predicted by the Q2D simulations is approximately 3 MHz, with less than 2 % error across all numbers of regions. This value closely matches the theoretical electron plasma oscillation frequency at the initial electron density, which is $\omega_{pe} \approx 2.8\ MHz$. Additionally, the captured damping rates of the oscillations are consistent with those from the full-2D simulation, with a maximum error of 9% in the simulation with the fewest regions ($M = N = 3$).

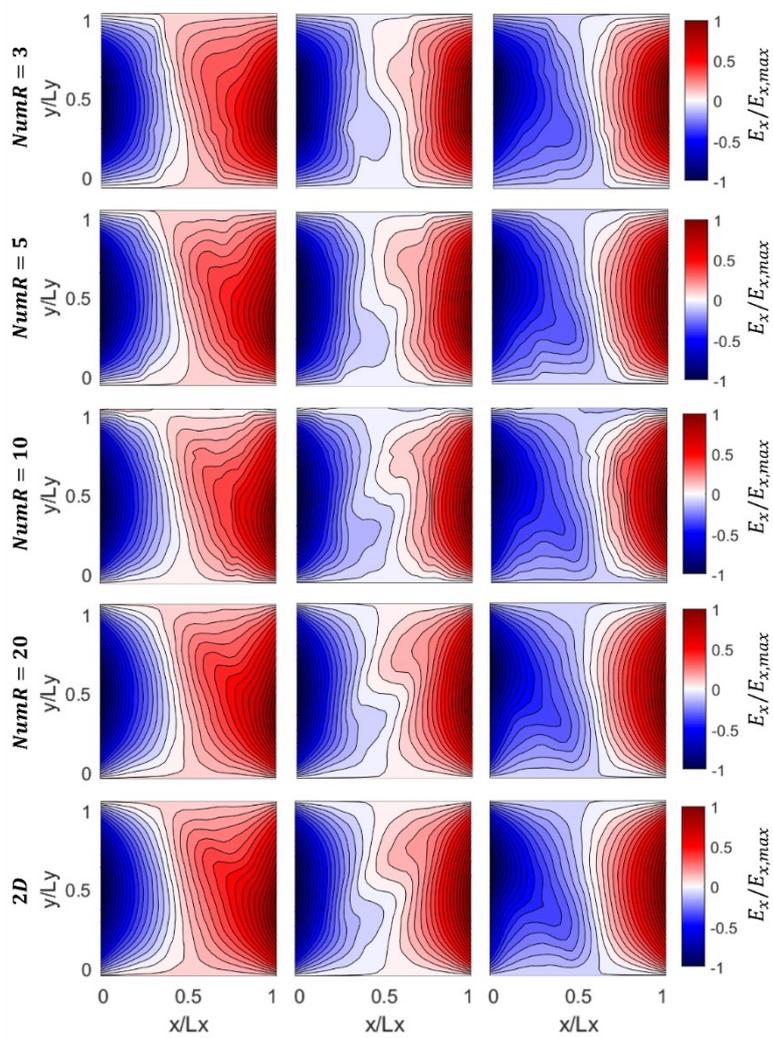

Figure 10: 2D snapshots of the normalized axial electric field ($E_x/E_{x,max}$, with $E_{x,max}$ being the peak $E_x$ in the respective snapshot from the full-2D simulation) from the first-order Q2D and the full-2D simulations at three different time instants of through the oscillations for test case 1.

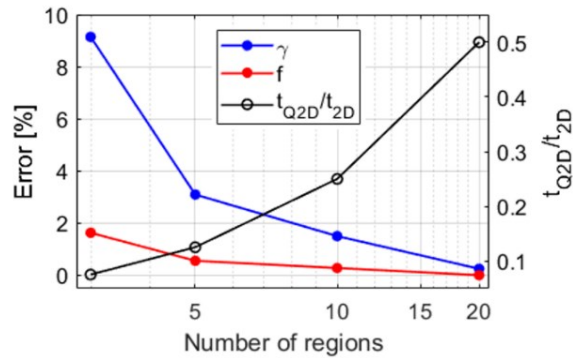

Figure 11: Variation of the error in predictions of the frequency (f) and the damping rate ($\gamma$) of the oscillations in the electric potential energy ($E_{pot}$) from the first-order Q2D simulations compared to full-2D results in test case 1. The computational time ratio between the Q2D and full-2D simulations are also plotted against the number of regions used for the Q2D simulations.

## 4.2. Test case 2: Diocotron instability problem

### 4.2.1. Description of the problem setup

This case captures the development of the Diocotron instability [38][39] with the overall setup of the simulation similar to those in Ref. [16].

Diocotron instability [38][39] typically occurs in plasmas confined within cylindrical or annular geometries and is characterized by the growth of azimuthal density perturbations. It results from the interplay between the radial electric field and the magnetic field applied parallel to the axis of the cylinder or annulus. The radial electric field causes the plasma to have varying rotational speeds across its radius, leading to sheared flow – a situation where different layers of the plasma rotate at different rates. This shearing effect amplifies the azimuthal disturbances and leads to the growth of instability modes into strong density variations and vortices. These evolving structures can significantly disrupt the plasma's intended confinement and stability.

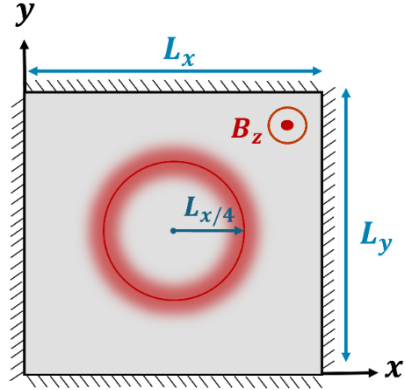

Figure 12: Schematic of the computational domain and setup of test case 2.

| Parameter | Value [unit] |
|---|---|
| Physical Parameters | |
| Initial electron temperature ($T_{e,0}$) | 1 [eV] |
| Background ion density | $1 \times 10^{12}$ [m$^{-3}$] |
| Applied magnetic field ($B_z$) | 2.5 [mT] |
| Potential at walls ($\phi_w$) | 0 [V] |
| Computational Parameters | |
| Time step ($\Delta t$) | $0.1\omega_{pe}$ |
| Time averaging for outputs | $\omega_{pe}$ |
| Total simulation time ($t_{sim}$) | $100\omega_{pe}$ |
| Domain length ($L_x = L_y$) | $25\lambda_D$ |
| Cell size ($\Delta x = \Delta y$) | $0.1\lambda_D$ |
| Number of computational nodes along each direction ($N_i = N_j$) | 250 |
| Initial number of macroparticles per cells ($N_{ppc}$) | 100 |
| Initial maximum plasma frequency ($\omega_{pe}$) | $5.64 \times 10^7$ [$rad/s$] |
| Initial minimum Debye length ($\lambda_D$) | $7.43 \times 10^{-1}$ [$cm$] |

Table 3: Summary of the main computational and physical parameters used for simulations of test case 2.

The simulation case of the Diocotron instability represents a 2D ($x - y$) Cartesian domain of length $L_x = L_y = 25\lambda_D$, where $\lambda_D$ denotes the Debye length. A schematic of the simulation domain is illustrated in Figure 12.

At the beginning of the simulation, electrons are sampled from a Maxwellian distribution with an initial temperature of $T_{e,0} = 1\ eV$. These electrons are loaded in space according to a hollow Gaussian distribution with a radius $R_{load} = L_x/4$ centered at $x = L_x/2$ and $y = L_y/2$ and a standard deviation of $\sigma = 0.03L_x$. A small perturbation is applied to the azimuthal position of the electrons on the Gaussian ring to expedite the formation of the instability. The ions form a uniform and immobile background with a density of $n_i = 1 \times 10^{12}\ m^{-3}$. A uniform magnetic field with an intensity of $B_z = 2.5\ mT$ is imposed perpendicular to the simulation plane. All domain boundaries are grounded with zero potential and a reflective boundary condition is applied to particles crossing the boundaries. The numerical and physical parameters of the simulation are summarized in Table 3.

*4.2.2. Results*

The results of the first-order Q2D simulations of test case 2 are presented in terms of snapshots of electron number density at different instances of the instability development in Figure 13, as well as the time-averaged plasma profiles along the $x$-axis at $y = 0$ in Figure 14. For comparison, the plasma profiles obtained from the zeroth-order Q2D simulations of test case 2 are also presented in Figure 14.

Additionally, Appendix B provides the corresponding snapshots of the electric potential from both the first-order and zeroth-order RO-PIC simulations, along with the snapshots of the electron number density from the zeroth-order RO-PIC.

The presented snapshots in Figure 13, as well as those in Appendix B, facilitate a direct comparison of how effectively each approximation degree in the Q2D simulations captures the evolving dynamics of the instability. The time-averaged plasma distributions in Figure 14 serve as a more quantitative comparison between the outcomes of the simulations.

Figure 13 and Figure 14 show that the first-order Q2D simulations with 25 and 50 regions closely reproduce the outcomes derived from the full-2D simulation. More specifically, the snapshots in Figure 13 illustrate that the detailed structure and temporal evolution of the instability is accurately represented with these approximation orders. Furthermore, the time-averaged profiles from these simulations in Figure 14 are almost indistinguishable from the full-2D profiles. The close agreement between the time evolution of the total electrons' kinetic energy and the electric potential energy from the 25- and 50-region simulations and from the full-2D simulation, provided in Appendix B, further supports the faithful representation of the underlying physics at these approximation levels.

With fewer regions (10 regions), the first-order Q2D simulation still manage to resolve the fundamental characteristics of the oscillations but exhibit subtle differences in finer details. The plasma profiles closely approximate the full-2D results.

Conversely, the Q2D simulation employing 4 regions shows greater discrepancies in the resolved structure of the instability. However, it provides an overall qualitative understanding of the behaviors and the processes to be expected. This is further evidenced from the time-averaged plasma profiles from the 4-region first-order Q2D simulation in Figure 14, which are not far off from those obtained from the full-2D simulations.

This variation in the resolution accuracy underscores a fundamental trade-off inherent in reduced-order modeling, i.e. the need to balance computational efficiency against results accuracy. While using fewer regions leads to a great reduction in computational load, it may also risk oversimplifying the complex dynamics involved, potentially leaving out key physical characteristics of the system. It is important to note that this trade-off must be considered in the context of the simulation's objectives. When high accuracies are not essential, a simplified representation provided by faster approximations can be sufficient. This is particularly the case when investigating the relative impact of a parameter or conducting an extensive parametric study to obtain an overview of the plasma system's behavior across a vast parameter space, which may be impractical or unnecessary with higher-accuracy approximations. Another example is during the preliminary design optimization phase of a plasma technology, where a detailed understanding of the underlying physics may not be critical. On the other hand, when the aim is a rigorous physics study of a phenomena, opting for higher approximation accuracy provided by the RO-PIC becomes essential to capture the intricacies of the involved dynamics accurately, while still gaining significant speedups relative to conventional PICs.

Now, we would spend some words discussing the relative performance of the first-order and the zeroth-order RO-PIC. Comparing the time-averaged profiles obtained from the zeroth-order and first-order simulations in Figure 14, along with the snapshots in Figure 13 and those presented in Appendix B from the zeroth-order version, reveals that the first-order formulation leads to marked improvement in the result's accuracy, particularly at lower number of regions. However, it is worth noting that the zeroth-order Q2D simulations with higher number-of-regions (50 regions, e.g.) still show good agreement with the full-2D simulation.

The improvements in the accuracy of the entire RO-PIC scheme in this plasma test case when adopting the first-order formulation is consistent with the differences that we observed in Section 3 between the zeroth-order and the first-order solutions from the RDPS for standalone Poisson problems.

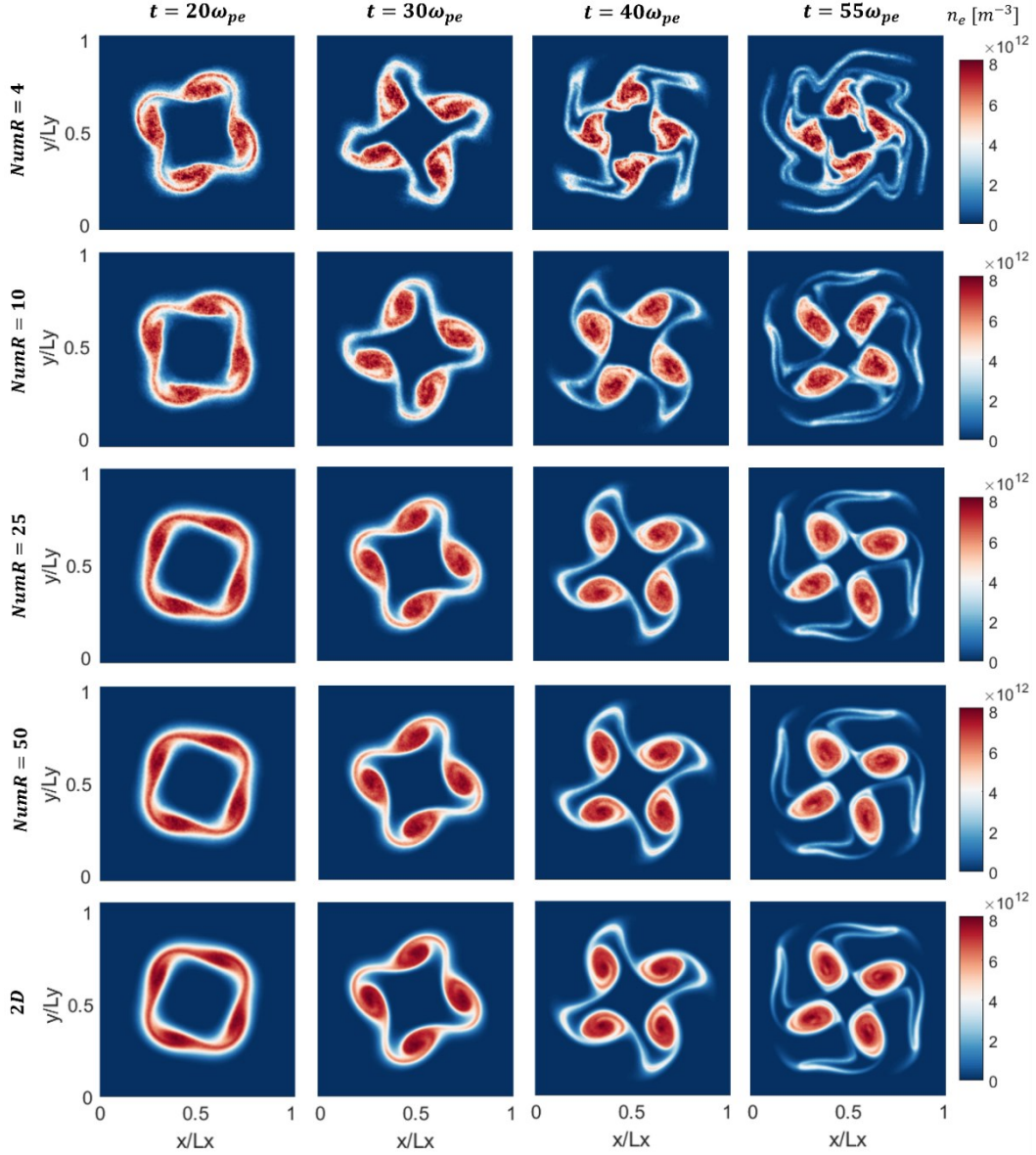

Figure 13: 2D snapshots of the electron number density ($n_e$) from the first-order Q2D and the full-2D simulations at four different time instants of the Diocotron instability evolution.

To conclude the discussion on the comparison between the accuracy of the zeroth-order and the first-order schemes, as well as the trade-off between the accuracy and computational cost, we refer to Figure 15.

This figure provides the convergence of the L2 errors in predicting various plasma properties from both schemes vs the number of regions used. The L2 error quantifies the average normalized difference between the time-averaged plasma profiles obtained from the Q2D and full-2D simulations, calculated on a node-by-node basis.

Expectedly, the errors decrease with increasing number of regions in both the zeroth-order and the first-order Q2D simulations. However, not only does the first-order scheme exhibit a significantly steeper rate of error reduction, but also the absolute error values are consistently lower compared to those in the zeroth-order scheme.

The higher error levels observed in the zeroth-order simulations can be of course attributed, in part, to a slight misalignment between the predicted Q2D profiles and the corresponding 2D profiles, as was seen in Figure 14. In any case, the zeroth-order simulations with 25 and 50 regions exhibit errors below 5% across all relevant plasma profiles, while errors in the first-order simulations with the same number-of-regions remain under 0.05% for all the plasma profiles except for the electron temperature profile, which has an error of about 2-3%. It is noted that these accuracies are achieved with simulations that have only 10-20 % of the computational time of the full-2D.

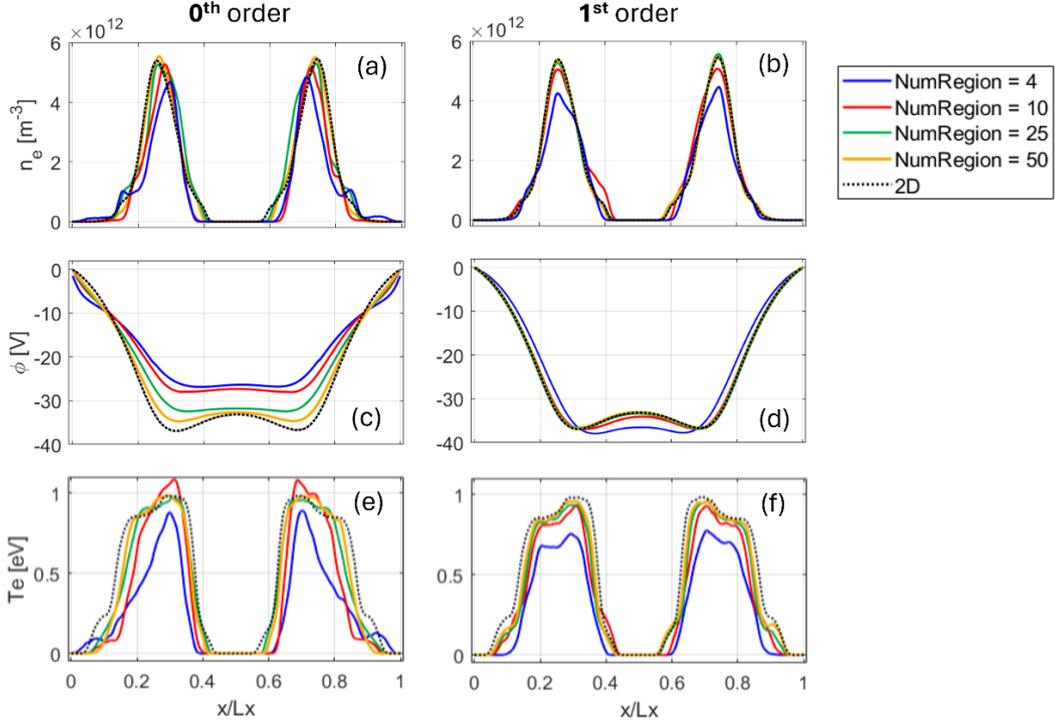

Figure 14: Time-averaged (over $0 - 60\omega_{pe}$) profiles of several plasma properties for test case 2 along the $x$-direction at the mid-plane ($y = L_y/2$) of the domain, from the zeroth-order (**left column**) and the first-order (**right column**) Q2D simulations with various number-of-regions, and from the full-2D simulation; (a) and (b) electron number density ($n_e$), (c) and (d) plasma potential ($\phi$), and (e) and (f) electron temperature ($T_e$).

In the first-order simulations, even with as few as 4 regions, the errors in both the electron number density ($n_e$) and the electric potential ($\phi$) distributions remain within 5 %. Additionally, the electron temperature ($T_e$) profile, has an error of approximately 15%. These error levels could be particularly acceptable for some purposes, especially given that they are obtained with only 1.6 % of the computational time of a full-2D simulation.

The electron temperature profiles were observed to generally exhibit higher error levels. To further investigate this, we conducted Q2D simulations with an increased number of macroparticles per cell, with the associated results discussed in subsection 4.2.2.1.

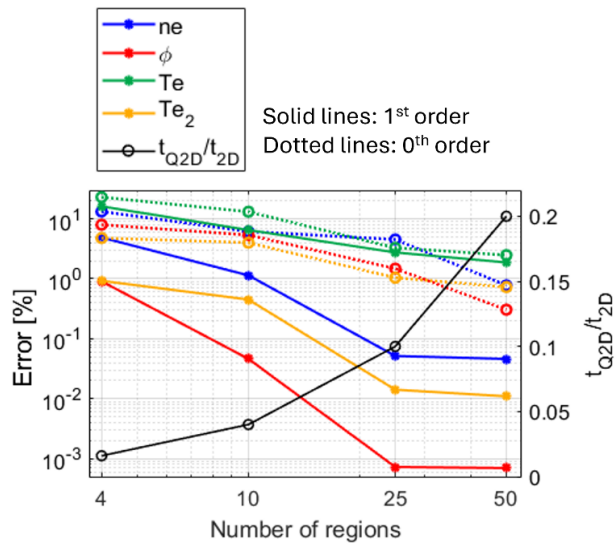

Figure 15: Variation of the L2 error in predictions of plasma profiles from the zeroth-order (dotted lines) and first-order (solid lines) Q2D simulations compared to full-2D results in test case 2. The computational time ratio between the Q2D and full-2D simulations are also plotted against the number of regions in the Q2D simulations. The yellow lines represent the errors in electron temperatures ($Te_2$) in the Q2D simulations with the macroparticle counts equivalent to those in the full-2D.

*4.2.2.1. Analysis of sensitivity to the number of macroparticles per cell*

As pointed out above, predicting the electron temperature accurately appeared more challenging than estimating quantities like density or potential. The reason is that the electron temperature is directly related to the average thermal energy, which depends on the square of particles' velocities. The particles' velocity distribution is highly sensitive to statistical noise, and when under sampled, it can lead to non-negligible errors in temperature calculations. This noise results from the limited number of macroparticles used in the simulations, which may introduce fluctuations in the velocity moments of the distribution functions deposited on the computational grid. Increasing the number of macroparticles per cell reduces this noise, yielding a more statistically representative velocity distribution and, consequently, a more precise electron temperature estimation.

With the above in mind, to assess the sensitivity of the Q2D predictions to the number of macroparticles per cell, we repeated the reduced-order simulations of test case 2 using a higher number of macroparticles per cell. For these new simulations, the total number of macroparticles matched that in the full-2D simulation, regardless of the number-of-regions used.

The comparison between the time-averaged plasma profiles obtained from the Q2D simulations using the original macroparticle counts (corresponding to the respective Q2D approximations) and those with a higher macroparticle count (corresponding to full-2D simulation) is presented in Figure 16.

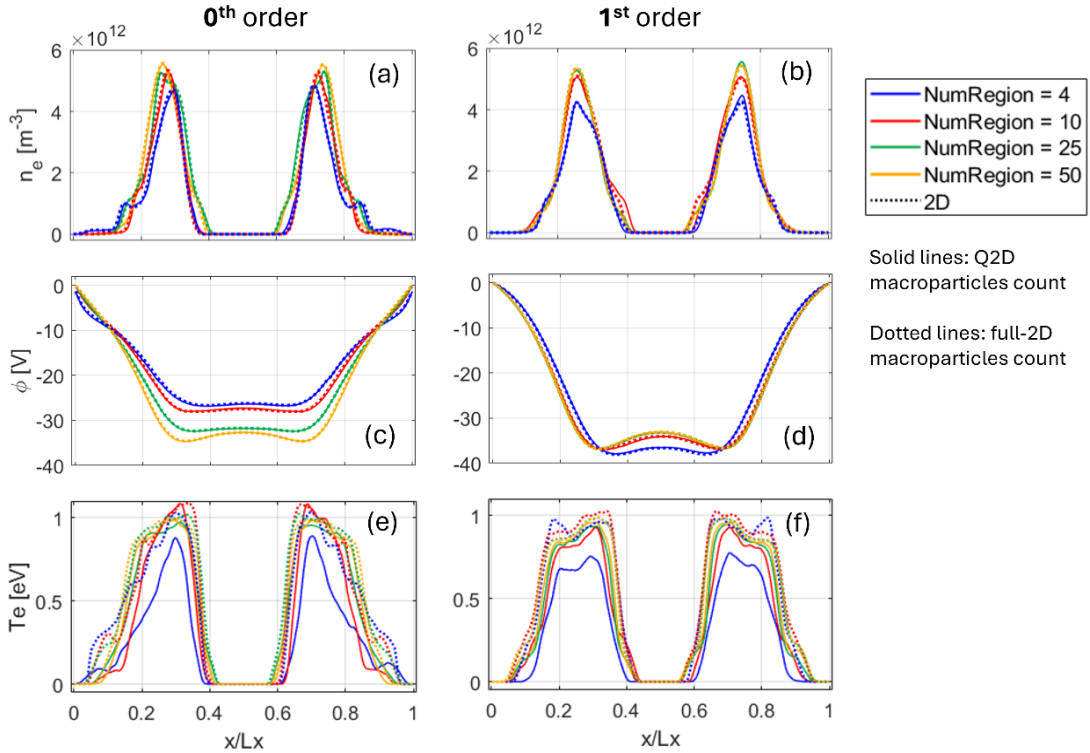

Figure 16: Comparison of the time-averaged (over $0 - 60\omega_{pe}$) profiles of several plasma properties for test case 2 along the $x$-direction at the mid-plane ($y = L_y/2$) of the domain from the Q2D simulations with the Q2D macroparticle counts (solid lines) and full-2D macroparticle counts (dotted lines); (**left column**) results from the zeroth-order and (**right column**) from the first-order Q2D simulations with various number of regions; (a) and (b) electron number density ($n_e$), (c) and (d) plasma potential ($\phi$), and (e) and (f) electron temperature ($T_e$).

The comparison using Figure 16 reveals that increasing the number of macroparticles does not appreciably impact the electron number density and the plasma potential profiles, as both remain consistent across different macroparticle counts.

However, the electron temperature profiles from the low-number-of-region simulations show notable improvement with higher macroparticle counts as more statistically representative particle velocity distributions now lead to more accurate representations of the electron temperature. This observation aligns with the earlier discussion on the sensitivity of the electron temperature to macroparticle sampling.

It is noted that the total macroparticle count in the original Q2D simulations, with higher number of regions, was already relatively close to that of the full-2D simulations. As a result, further increasing the macroparticle count has a limited impact on the temperature profiles for those approximation degrees.

The errors in the temperature profiles from various Q2D simulations with higher macroparticle counts were presented in Figure 15. Additionally, Figure 17 compares several electron density snapshots from a first-order Q2D simulation with 4 regions, with two different macroparticle counts. This comparison further demonstrates that increasing the number of macroparticles does not significantly alter the density profile; rather, it enhances the quality and resolution of the velocity distribution function, leading to improved estimations of the temperature.

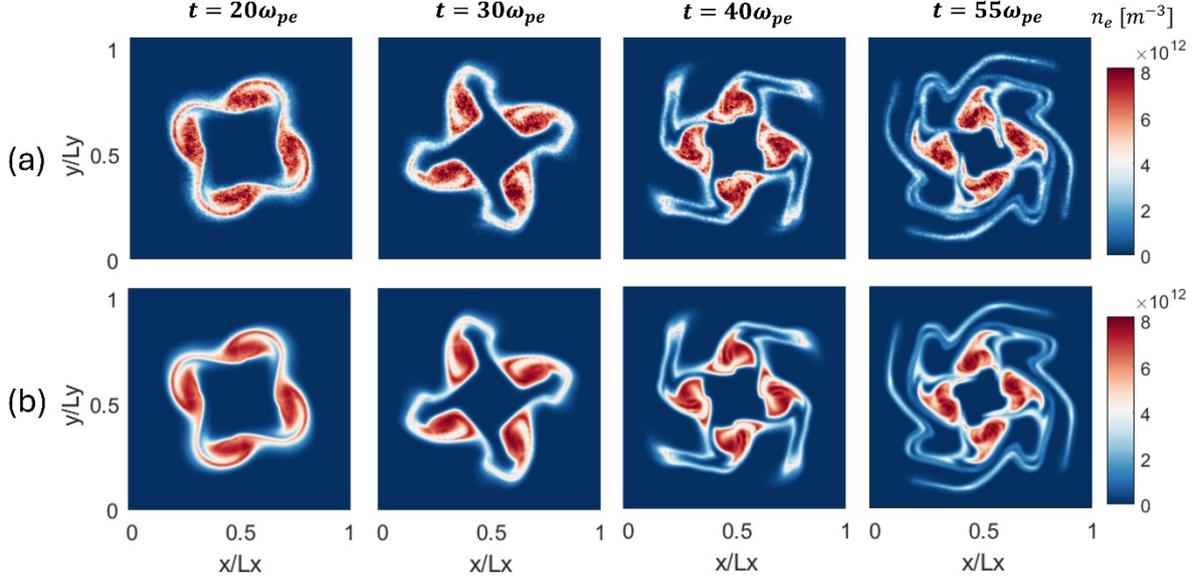

Figure 17: Comparison of the 2D snapshots of the electron number density ($n_e$) from the first-order Q2D simulations with number of regions $NumR = 4$, and with (a) the Q2D macroparticle count and (b) the full-2D macroparticle count. The snapshots correspond to four different instants of time during the Diocotron instability evolution.

**Section 5: Conclusions**

This paper presented advancements in the reduced-order PIC scheme, focusing on the development of the first-order Q2D formulation. Compared to the original zeroth-order version published in 2023 [18], the first-order approach significantly improved computational efficiency and accuracy. These were demonstrated through verifications of the RDPS module within RO-PIC in standalone Poisson problems, as well as the complete RO-PIC code in representative plasma test cases involving electrons plasma oscillations with Landau damping and Diocotron instability.

The first-order Q2D formulation introduces a linear variation of data across regions, unlike the zeroth-order formulation, which assumes data to be constant within each region. This distinction enables the first-order Q2D to achieve greater accuracy and smoothness while requiring fewer computational resources. The verifications of the upgraded RDPS showed that the first-order formulation can match the full-2D solution accuracy with far less computational cells, making it suitable for integration within large-scale plasma simulations.

Following this, in the plasma test cases, the first-order Q2D simulations with higher number of regions accurately captured key plasma behaviors, with results closely matching those from the full-2D simulations while offering up to 10 times speedup compared to a standard 2D PIC. Even with fewer regions, the Q2D simulations effectively resolved essential dynamics, balancing accuracy and computational efficiency.

The first-order formulation consistently showed improved accuracy compared to the zeroth-order one, especially at lower approximation levels to the full-2D problems. Furthermore, faster convergence rates were observed in the first-order simulations, which suggests that, as the number of regions continues to increase, the method reach near-exact accuracy with significantly fewer regions compared to the what was the case with the zeroth-order approach.

In summary, the first-order RO-PIC offers an effective solution to the computational challenges of traditional PIC simulations, enabling quasi-multi-dimensional approximations with flexible trade-offs between cost and accuracy. This method is both efficient and versatile, making self-consistent plasma modelling more practical for a wide range of scientific and applied applications. In this regard, the range of approximations offered by the RO-PIC provides users with a unique flexibility to choose the right balance between computational cost and fidelity based on their simulation needs. Low number-of-regions are ideal for rapid exploratory studies and preliminary designs, while higher number-of-regions deliver the high accuracy needed for detailed physics studies, still providing substantial computational savings compared to traditional PIC methods.

Having introduced the first-order RO-PIC implementation in a 2D configuration in this part I, and encouraged by the extensive verification results, we extend the first-order formulation to 3D configuration in part II and examine in detail our resulting Q3D first-order RO-PIC code.

# Appendix

## A. Additional Poisson problems

Completing the test cases presented in Section 3 for standalone verifications of the first-order RDPS, we provide a set of additional cases in this appendix, which are detailed in Table 4.

In this regard, unlike cases 1 and 2 in Section 3 and 3 to 6 here, which all involve Dirichlet boundary conditions, case 7 features periodic/Neumann condition along the $y$-direction.

| Case No. | Source terms ($\rho(x,y)$) | Boundary Conditions |
|---|---|---|
| 3 | $\rho = 0$ | $\phi(0,y) = \dfrac{y}{1+y^2}$; $\phi(L_x, y) = \dfrac{y}{4+y^2}$ <br> $\phi(x, 0) = 0$; $\phi(x, L_y) = \dfrac{1}{(1+x)^2 + 1}$ |
| 4 | $\rho = x^2 + y^2$ | $\phi(0,y) = \sin(\pi y)$ <br> $\phi(L_x, y) = \exp(\pi L_x)\sin(\pi y) + 0.5 y^2$ <br> $\phi(x, 0) = 0$; $\phi(x, L_y) = 0.5 x^2$ |
| 5 | $\rho = \exp(xy)$ | $\phi(0,y) = 5(y^2 - y)$; $\phi(L_x, y) = -y(y-1)^4$ <br> $\phi(x, 0) = 0.5\sin(6\pi x)$; $\phi(x, L_y) = \sin(2\pi x)$ |
| 6 | $\rho = 100(x - 0.5)^3$ | $\phi(0,y) = 0$; $\phi(L_x, y) = 0$ <br> $\phi(x, 0) = -(0.5)^3$; $\phi(x, L_y) = (0.5)^3$ |
| 7 | $\rho = y\sin(5\pi x) + \exp(-50((x-0.5)^2 + (y-0.5)^2))$ | $\phi(0,y) = 0$; $\phi(L_x, y) = 0$ <br> At $y = 0, L_y$ : Periodic B.C./ Neumann B.C. |

Table 4: Definition of additional 2D Poisson problems for standalone verifications of the first-order RDPS.

The contour maps of the solutions from the first-order and the zeroth-order RDPS with various number of regions are presented in Figure 18 for cases 3 and 4, and in Figure 19 for case 5 and 6.

These figures again illustrate the enhanced accuracy and smoothness of the first-order solutions compared to the zeroth-order results, particularly at low numbers of regions. Indeed, the approximate first-order solutions obtained using only a few regions ($M, N = 3, 5$) are seen to qualitatively well resemble the full-2D solution.

Figure 20 shows the solutions of the first-order RDPS for case 7 across different numbers of regions. As evident from the depicted contour plots, the solutions exhibit seamless periodicity along the $y$-axis in Figure 20(a) and zero gradient (consistent with the Neumann condition) in Figure 20(b).

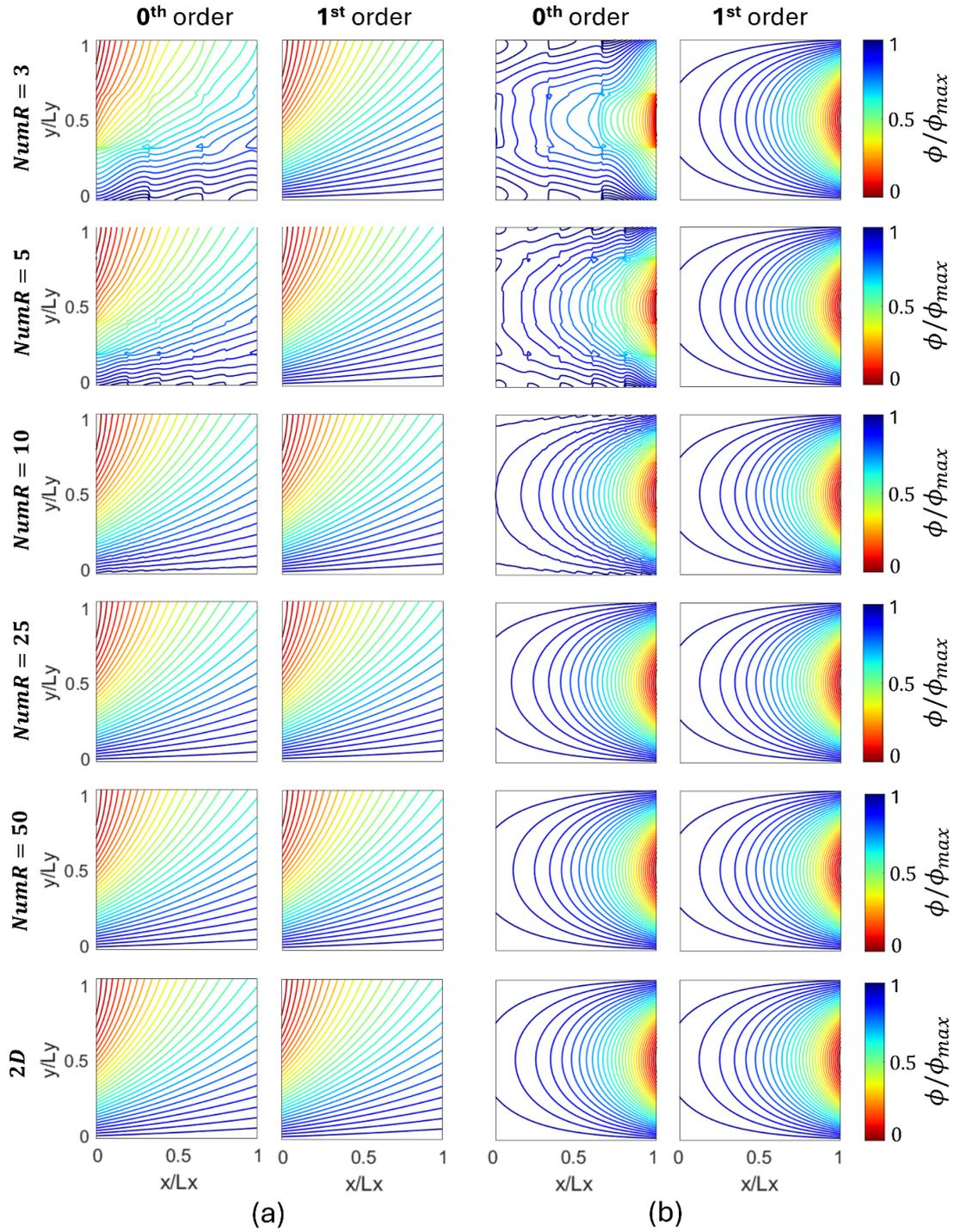

Figure 18: Comparison of the zeroth-order and the first-order RDPS' solutions with various number of regions against the corresponding full-2D solution (**bottom-most row**) for (a) case 3 and (b) case 4. The numbers of regions are equal along both dimensions ($M = N$), which are denoted by *NumR*.

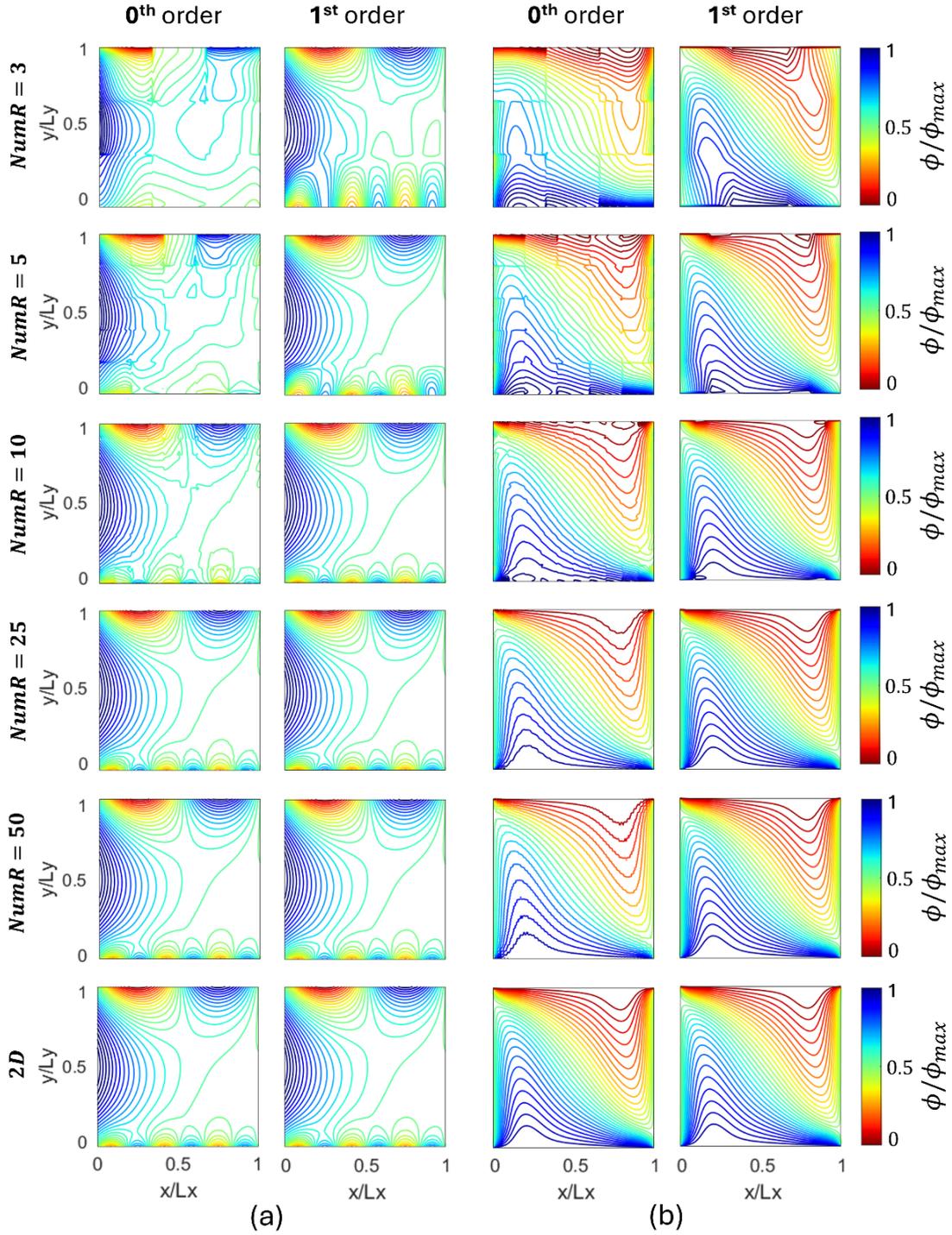

Figure 19: Comparison of the zeroth-order and the first-order RDPS' solutions with various number of regions against the corresponding full-2D solution (**bottom-most row**) for (a) case 5 and (b) case 6. The numbers of regions are equal along both dimensions ($M = N$), which are denoted by *NumR*.

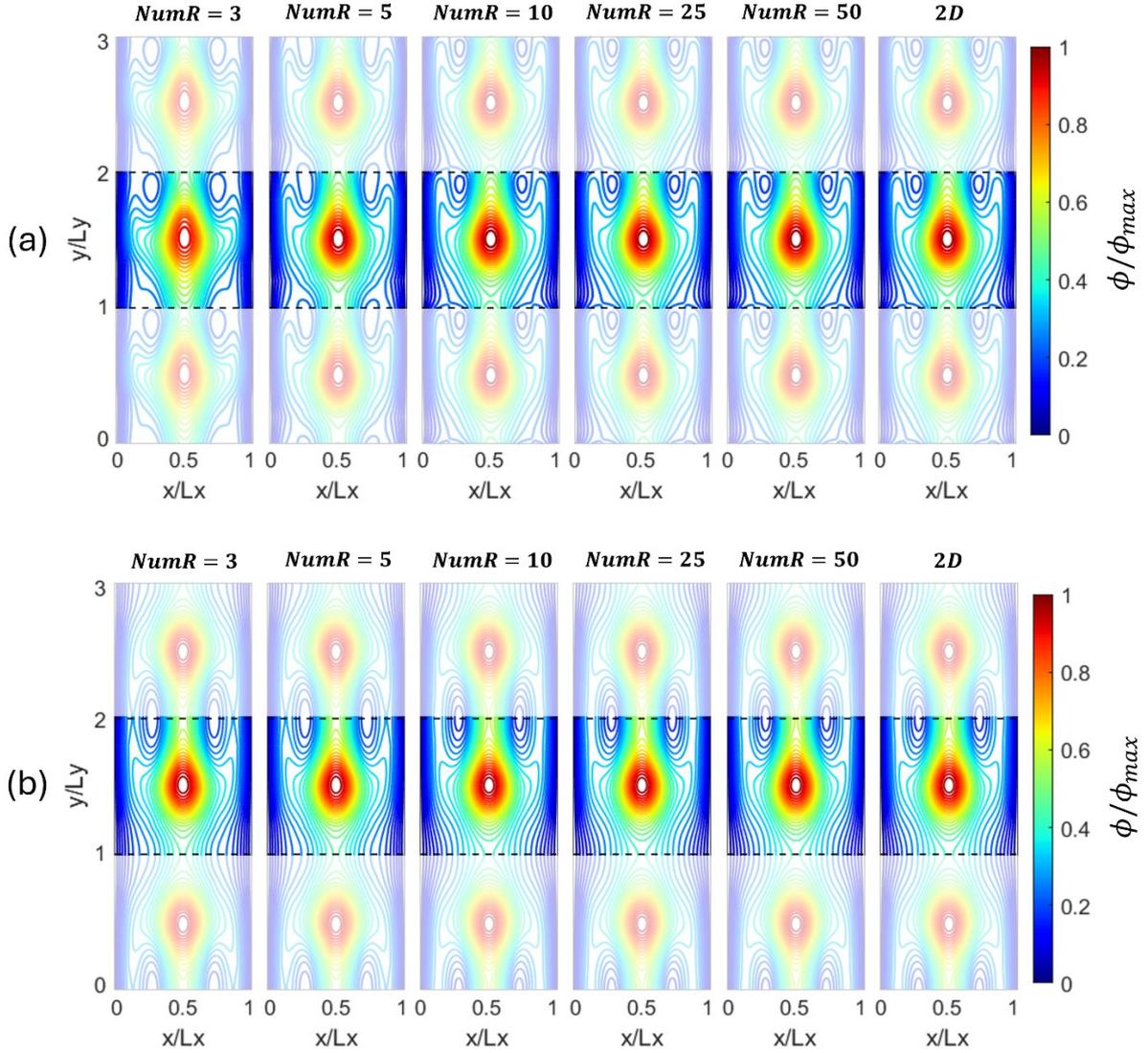

Figure 20: Comparison of the first-order RDPS' solutions with various number of regions against the corresponding full-2D solution (**rightmost column**) for case 7 with (a) periodic boundary condition and (b) Neumann boundary condition along the $y$-direction. The numbers of regions are equal along both dimensions ($M = N$), which are denoted by $NumR$. Potential is solved only for $y/L_y \in [0,1]$ and is duplicated twice along the $y$-direction to illustrate the satisfaction of periodicity in (a) and the zero-gradient condition in (b) for the respective cases.

## B. Additional results for the Diocotron instability test case

This appendix contains additional results from the Q2D simulations of the Diocotron instability case, providing further support for the arguments presented in the main text within subsection 4.2.2.

In particular, supplementary results from the first-order Q2D simulations for this test case are illustrated in Figure 21, which displays 2D snapshots of the electric potential at different moments during the evolution of the instability. Additionally, Figure 22 shows the time variations of the total kinetic energy of the electrons and the electric potential energy from the first-order Q2D simulations with 25 and 50 regions.

The results from the zeroth-order Q2D simulations of the Diocotron instability case are provided in Figure 23 and Figure 24, showing, respectively, the snapshots of the electron number density and the electric potential at four distinct time instants during the development of the instability. These figures, when compared against the snapshots obtained from the first-order Q2D simulations, highlight the effectiveness of the first-order formulation in enhancing the prediction accuracy of the RO-PIC.

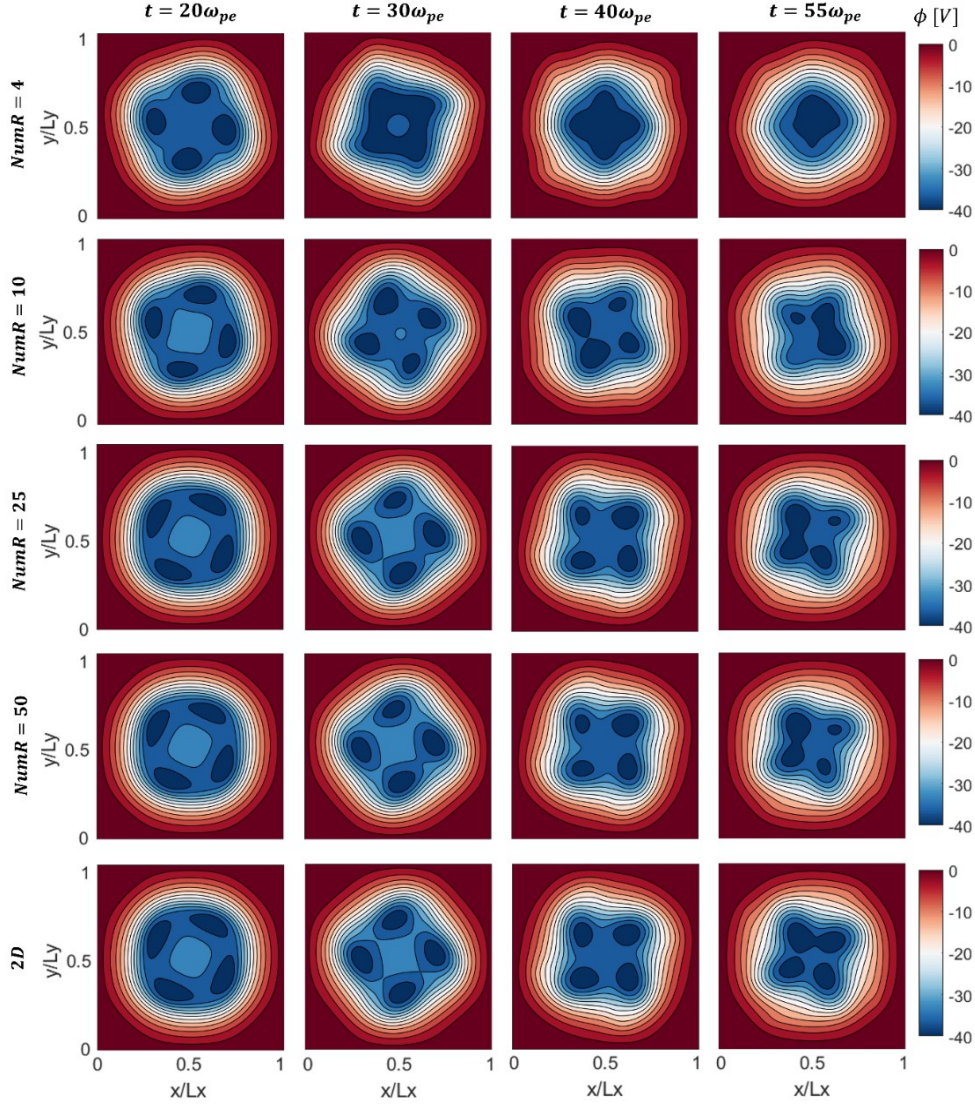

Figure 21: 2D snapshots of the electric potential ($\phi$) from the first-order Q2D and the full-2D simulations at four different time instants during the Diocotron instability evolution.

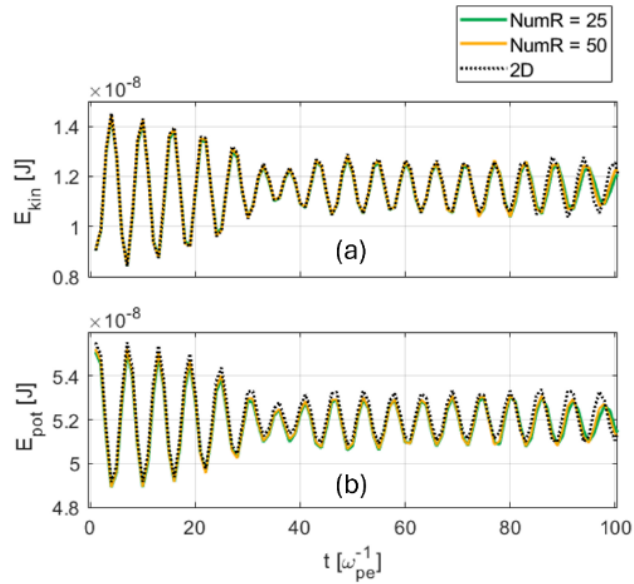

Figure 22: Time evolution of (a) total electrons' kinetic energy ($E_{kin}$) and (b) electric potential energy ($E_{pot}$), from the Q2D (with 25 and 50 regions) simulations and the full-2D simulations for the Diocotron instability case.

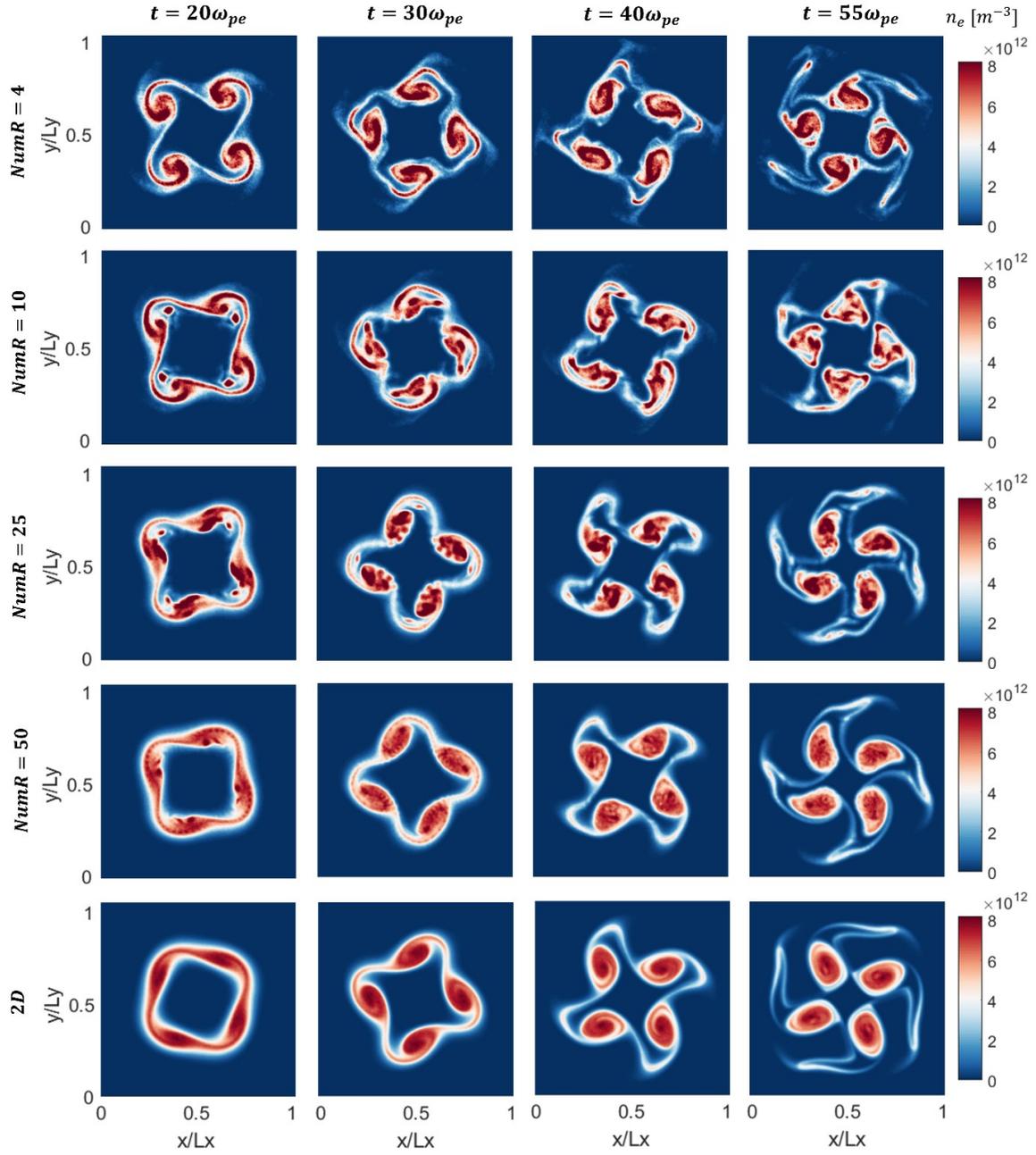

Figure 23: 2D snapshots of the electron number density ($n_e$) from the zeroth-order Q2D and the full-2D simulations at four different time instants during the Diocotron instability evolution.

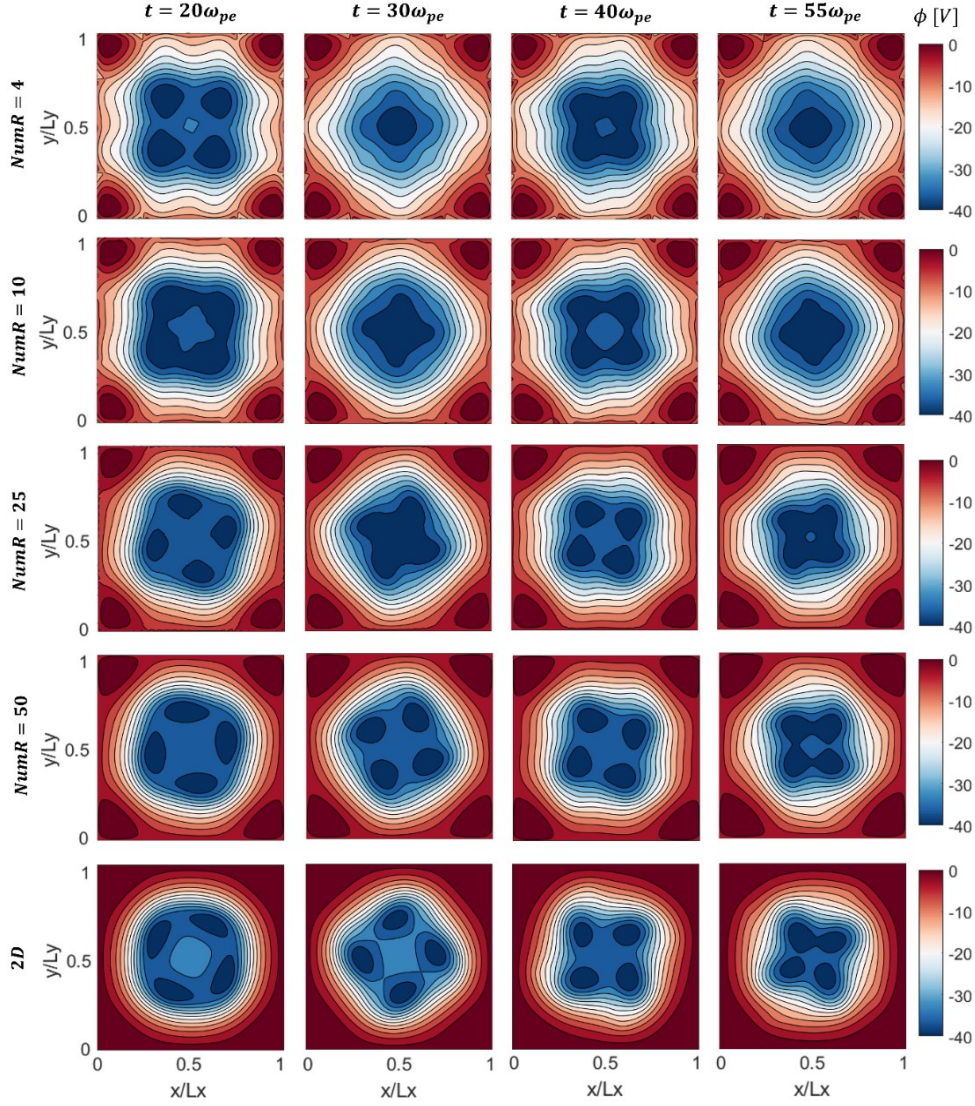

Figure 24: 2D snapshots of the electric potential ($\phi$) from the zeroth-order Q2D and the full-2D simulations at four different time instants during the Diocotron instability evolution.

**References**:

[1] Lafleur T, Baalrud SD, Chabert P, "Theory for the anomalous electron transport in Hall effect thrusters. I. Insights from particle-in-cell simulations," *Phys. Plasmas* **23**, 053502 (2016)

[2] Reza M, Faraji F, Andreussi T, Andrenucci M, "A Model for Turbulence-Induced Electron Transport in Hall Thrusters," IEPC-2017-367, *35th International Electric Propulsion Conference*, Atlanta, Georgia (2017)

[3] Marks TA, Jorns BA, "Evaluation of several first-principles closure models for Hall thruster anomalous transport," *In Proceedings of AIAA SciTech Forum 2023*, National Harbor, MD & Online (2023)

[4] Katz I, Chaplin VH, Ortega AL, "Particle-in-cell simulations of Hall thruster acceleration and near plume regions," *Phys. Plasmas* **25**, 123504 (2018)

[5] Mikellides IG, Jorns B, Katz I, Ortega AL, "Hall2De Simulations with a First-principles Electron Transport Model Based on the Electron Cyclotron Drift Instability," *52nd AIAA/SAE/ASEE Joint Propulsion Conference*, Salt Lake City, Utah (2016) DOI: 10.2514/6.2016-4618

[6] Mikellides IG, Lopez Ortega A, "Challenges in the development and verification of first-principles models in Hall-effect thruster simulations that are based on anomalous resistivity and generalized Ohm's law," *Plasma Sources Sci. Technol.* **28**, 014003 (2019)

[7] Mikellides IG, Lopez Ortega A, Chaplin VH, "Theory of the anomalous momentum exchange from wave-particle interactions in Hall-effect ion accelerators and comparisons with measurements," *Phys. Fluids* **36**, 074121 (2024)

[8] Lewis HR, "Energy-conserving numerical approximations for Vlasov plasmas," *Journal of Computational Physics*, Vol. **6**, Issue 1, pg. 136-141 (1970)

[9] Cohen BI, Langdon AB, Friedman A, "Implicit time integration for plasma simulation," *Journal of Computational Physics*, Vol. **46**, Issue 1, pg. 15-38 (1982)

[10] Barnes DC, Chacón L, "Finite spatial-grid effects in energy-conserving particle-in-cell algorithms," *Journal of Computational Physics*, Vol. **258**, 107560 (2021)

[11] Sun H, Banerjee S, Sharma S, Powis AT, Khrabrov AV, Sydorenko D, Chen J, Kaganovich I, "Direct implicit and explicit energy-conserving particle-in-cell methods for modeling of capacitively coupled plasma devices," *Phys. Plasmas* **30** (10): 103509 (2023)

[12] Cohen BI, Langdon AB, Hewett DW, Procassini RJ, "Performance and optimization of direct implicit particle simulation," *Journal of Computational Physics*, Vol. **81**, Issue 1, pg. 151-168 (1989)

[13] Zenger C, "Sparse Grids," In: W. Hackbusch, Ed., Notes on Numerical Fluid Mechanics, Vol. 31, Vieweg, Braunschweig, pp. 241-251 (1991)

[14] Griebel M, Schneider M, Zenger C, "A combination technique for the solution of sparse grid problems," Technische Universität (1990)

[15] Ricketson LF, Cerfon AJ, "Sparse grid techniques for particle-in-cell schemes," *Plasma Phys. Control. Fusion* **59**, 024002 (2016)

[16] Deluzet F, Fubiani G, Garrigues L, Guillet C, Narski J, "Sparse grid reconstructions for particle-in-cell methods," *ESAIM: M2AN* **56**, 1809–1841 (2022)

[17] Garrigues L, Chung-To-Sang M, Fubiani G, Guillet C, Deluzet F, Narski J, "Acceleration of particle-in-cell simulations using sparse grid algorithms. II. Application to partially magnetized low temperature plasmas," *Phys. Plasmas* **31**, 073908 (2024)

[18] Reza M, Faraji F, Knoll A, "Concept of the generalized reduced-order particle-in-cell scheme and verification in an axial-azimuthal Hall thruster configuration", *J. Phys. D: Appl. Phys.* **56** 175201 (2023)

[19] Reza M, Faraji F, Knoll A, "Generalized reduced-order particle-in-cell scheme for Hall thruster modeling: concept and in-depth verification in the axial-azimuthal configuration", ArXiv, arXiv:2208.13106 (2022)

[20] Faraji F, Reza M, Knoll A, "Verification of the generalized reduced-order particle-in-cell scheme in a radial-azimuthal E×B plasma configuration", *AIP Advances* **13**, 025315 (2023)

[21] Faraji F, Reza M, Knoll A, "Enhancing one-dimensional particle-in-cell simulations to self-consistently resolve instability-induced electron transport in Hall thrusters". *J. Appl. Phys.* **131**, 193302 (2022)

[22] Reza M, Faraji F, Knoll A, "Resolving multi-dimensional plasma phenomena in Hall thrusters using the reduced-order particle-in-cell scheme", *J Electr Propuls* **1**, 19 (2022)

[23] Reza M, Faraji F, Knoll A, "Latest verifications of the reduced-order particle-in-cell scheme: Penning discharge and axial-radial Hall thruster case", AIAA 2024-2712, In Proceedings of 2024 SciTech Forum conference, Orlando, Florida (2024)

[24] Charoy T, Boeuf JP, Bourdon A, Carlsson JA, Chabert P, Cuenot B, Eremin D et al., "2D axial-azimuthal particle-in-cell benchmark for low-temperature partially magnetized plasmas," *Plasma Sources Sci. Technol.* **28**, 105010 (2019)

[25] Villafana W, Petronio F, Denig AC, Jimenez MJ, Eremin D, Garrigues L, Taccogna F, Alvarez-laguna A, Boeuf JP, Bourdon A, Chabert P, Charoy T, Cuenot B, Hara K, Pechereau F, Smolyakov A, Sydorenko D et al., "2D radial-azimuthal particle-in-cell benchmark for E×B discharges", *Plasma Sources Sci. Technol.* **30** 075002 (2021)

[26] See "https://jpb911.wixsite.com/landmark/copie-de-test-case-3-fluid-hybrid" for details of the Penning Discharge Benchmark activity (webpage last accessed in October 2024).

[27] Reza M, Faraji F, Knoll A, "Effects of the applied fields' strength on the plasma behavior and processes in E×B plasma discharges of various propellants: I. Electric field," *Phys. Plasmas.* **31**, 032120 (2024)

[28] Reza M, Faraji F, Knoll A, "Effects of the applied fields' strength on the plasma behavior and processes in E×B plasma discharges of various propellants: II. Magnetic field," *Phys. Plasmas.* **31**, 032121 (2024)

[29] Reza M, Faraji F, Knoll A, "Parametric investigation of azimuthal instabilities and electron transport in a radial-azimuthal E×B plasma configuration," *Journal of Applied Physics* **133**, 123301 (2023)

[30] Reza M, Faraji F, Knoll A, "Influence of the magnetic field curvature on the radial-azimuthal dynamics of a Hall thruster plasma discharge with different propellants," *J. Appl. Phys.* **134**, 233303 (2023)

[31] Reza M, Faraji F, Knoll A, "Plasma dynamics and electron transport in a Hall-thruster-representative configuration with various propellants: I. Variations with discharge voltage and current density," *Plasma* **7**(3), 651-679 (2024)

| | |
|---|---|
| [32] | Reza M, Faraji F, Knoll A, "Plasma dynamics and electron transport in a Hall-thruster-representative configuration with various propellants: II. Effects of the magnetic field topology," *Plasma* **7**(3), 680-704 (2024) |
| [33] | Reza M, Faraji F, Knoll A, Piragino A, Andreussi T, Misuri T, "Reduced-order particle-in-cell simulations of a high-power magnetically shielded Hall thruster," *Plasma Sources Sci. Technol.* **32**, 065016 (2023) |
| [34] | Birdsall CK, Langdon AB, "Plasma Physics via Computer Simulation", CRC Press, 1991 |
| [35] | Taccogna F, Cichocki F, Eremin D, Fubiani G, Garrigues L, "Plasma propulsion modeling with particle-based algorithms", *J. Appl. Phys.* **134**, 150901 (2023) |
| [36] | Reza M, Faraji F, Knoll A, "Latest progress on the reduced-order particle-in-cell scheme: II. Quasi-3D implementation and verification," Submitted manuscript, in review. |
| [37] | Brieda L, "Plasma Simulations by Example," CRC Press, 2019. |
| [38] | MacFarlane GC, Hay HG, "Wave Propagation in a Slipping Stream of Electrons: Small Amplitude Theory," *Proc. Roy. SOC.* **63**, B409 (1953) |
| [39] | Levy RH, "The diocotron instability in a cylindrical geometry," Airforce Office of Scientific Research Report no. 202, (1964) |
| [40] | Bittencourt JA, "Fundamentals of Plasma Physics," 3rd Edition, New York, NY: Springer; 2004. |